\documentclass[twocolumn,amsmath,amssymb]{revtex4-2}

\usepackage{graphicx}
\usepackage[utf8]{inputenc}
\usepackage{amsmath,amssymb}
\usepackage{epstopdf}
\usepackage{hyperref}
\usepackage{lineno}
\usepackage{color}
\usepackage[normalem]{ulem}

\newcommand{\Fref}[1]{Fig.~\ref{#1}}
\newcommand{\Eqref}[1]{Eq.~(\ref{#1})}
\newcommand{\be}{\begin{equation}}
\newcommand{\ee}{\end{equation}}
\newcommand{\bal}{\begin{align}}
\newcommand{\eal}{\end{align}}
\newcommand{\bear}{\begin{eqnarray}}
\newcommand{\eear}{\end{eqnarray}}
\newcommand{\nn}{\nonumber}

\newcommand{\md}{\mathrm{d}}
\newcommand{\e}{\mathrm{e}}
\newcommand{\cw}{C_\mathrm{W}}
\newcommand{\kb}{k_{_\mathrm{B}}}
\newcommand{\alf}{{Alfv\'en~}}
\newcommand{\im}{\mathrm{i}}

\newcommand{\AcousticHeating}{Gudiksen:05,ChromoView:19,Pelekhata:23,Vytenis:25,Udnaes:25}

\begin{document}

\title{On the Theory of Bulk Viscosity of Cold Plasmas and\\ 
Thermodynamics of Alkali-Noble Plasma Cocktails}

\author{Albert~M.~Varonov}
\email[Vice-corr. author: ]{akofmg@gmail.com}
\affiliation{Georgi Nadjakov Institute of Solid State Physics, Bulgarian Academy of Sciences,\\
72 Tzarigradsko Chaussee Blvd., BG-1784 Sofia, Bulgaria}

\author{Todor~M.~Mishonov}
\email[Corr. author: ]{mishonov@gmail.com}
\affiliation{Georgi Nadjakov Institute of Solid State Physics, Bulgarian Academy of Sciences,\\
72 Tzarigradsko Chaussee Blvd., BG-1784 Sofia, Bulgaria}

\date{7 Aug 2026}

\begin{abstract}
By solving the kinetic equation for ionization-recombination processes 
in cold plasmas for temperatures much lower than the first ionization potentials,
we derive an explicit expression for the complex polytropic index, as well as bulk viscosity.
The obtained result for the ionization-recombination relaxation time reveals that the Mandelstam-Leontovich approximation for the frequency dependence of the bulk viscosity is actually an exact solution in the case of cold plasmas.
We systemize also explicit general formulae for the thermodynamics of the alkali-noble cocktails in the same low temperature plasma approximation,
allowing the Schwarzschild criterion on convectional instability to be calculated for H-He and arbitrary plasma cocktail too.
All these general results are applicable for cold stellar, interstellar and laboratory plasmas.
Additionally, the wave heating of the inner solar atmosphere up to the solar transition region is studied thoroughly, and it is found that the magnetic diffusivity, heat conductivity, shear- and bulk viscosities all play an almost equally important part in the wave damping, therefore none of them can be omitted in realistic considerations of heating of the solar atmosphere.
\end{abstract}

\maketitle

\section{Introduction. Bulk viscosity in different physical conditions.}

Why do champagne glasses ring hollow during a toast?
What is the dominant mechanism for sound absorption in the ocean water?
Can we do a similar experiment in a glass of Epsom salts?
Are the models of the solar corona heating with acoustic waves realistic?
There is a huge number of physical systems in which bulk viscosity prevails over shear viscosity,
but due to the complexity of the problem, we are unable to calculate bulk viscosity from first principles.
Except perhaps, for the simplest case of cold plasma, which is the subject of our research.
Our purpose is to derive a convenient formula for the bulk viscosity applicable for the case when the temperature $T$ is much smaller than the ionization potential $I_a$ of the main atomic ingredients of a plasma cocktail.

\subsection{Motivation. Acoustic heating in the inner solar atmosphere}

In some sense our study is inspired by many articles where acoustic heating 
of the solar atmosphere is considered~\cite{\AcousticHeating}.
That is why for an illustration we give as a an example the solar chromosphere at 
height $h=1894$~km above the photosphere from the height dependent solar atmospheric profile~\cite[Model C7]{Avrett:08} (AL08).
In the solar plasmsa cocktail hydrogen dominates, therefore it is convenient to parameterize the chemical compound by the abundances of other elements $\overline a_a$ given in Ref.~\cite[Table~2]{Avrett:08}.
In such a way per every hydrogen atom we have $\mathcal{N}_a$ atoms
\be
\mathcal{N}_a=\sum_a \overline a_a \approx 1.1.
\ee
The concentrations of remaining chemical elements are correspondingly
$\overline a_a/\mathcal{N}_a$.
In such a way, 
if $n_\rho$ is the number of hydrogen atoms per unit volume,
the number of the other atoms per unit volume is
\be
n_a=n_\rho\,\overline a_a.
\ee

\subsection{A short discussion on the bulk viscosity}

In the air we breathe,
the energy transfer between translational and intramolecular degree of freedoms is relatively slow
and the first and second viscosity are comparable $\zeta/\eta\sim 1$.
Perhaps Albert Einstein was the first who paid attention on the theory
of bulk viscosity in partially dissociated gases~\cite{Einstein:20}.
The direct translation of this paper is \textit{Sound propagation in partially dissociated gases} and in this sense, our study is a continuation of Einstein's idea realized in partially ionized atomic gases with the main focus being the solar chromosphere.

Calculating the bulk viscosity from first principles has not been yet presented, as what is known to us so far.
In special cases, in the presence of strong magnetic fields, rotations~\cite{deGroot:54} and in molecular dynamics~\cite{LevanR:25,Levan:24,Levan:25} bulk viscosity analysis has been made, but in general ,the bulk viscosity is often neglected when treating plasma problems.

For the importance of the bulk viscosity $\zeta$ in the physics of plasmas 
we wish to point out the recent study by \citet{Istomin:17}
who studied transport coefficients and heat fluxes in non-equilibrium high-temperature flows with electronic excitation.
They confirm the intuitive conclusions by Einstein and calculate that 
chemical processes can lead to bulk viscosity domination $\zeta\gg\eta$.
The authors of this study perform a non-equilibrium kinetic calculation of N and O ionized mixtures, while our study is focused on monoatomic plasmas close to the local thermodynamic equilibrium (LTE).
Nevertheless, their conclusions and the results in our work are qualitatively the same.
Their conclusion is that the bulk viscosity coefficients are strongly affected by the electronic excitation which is quantitatively similar to ours for the influence of the ionization-recombination processes to the bulk viscosity and degree of ionization
$\alpha$.
\citet[Figure~3]{Istomin:17} in their study
demonstrates that the bulk viscosity has a broad maximum as a function of the temperature
and significantly exceeds the shear one analogously to the
atomic plasma which is the subject in the present study.
For the hydrogen-helium cocktail (H-He) in the present article we derive for the dispersionless bulk viscosity
$\zeta_0\propto (1-\alpha)\alpha$,
and at fixed mass density $\rho$,
the ionization degree $\alpha(T)$ monotonously increases between 0 and 1, and again as a function of the temperature we have a broad maximum.

\citet{Istomin:17} emphasize how important is for the calculation of the bulk viscosity is to have reliable data for the temperature dependence of the rates of the chemical reactions.
Our main motivation is that for a hydrogen-like atom we have a reliable formula for the near threshold ionization cross-section for more than 70 years.
which gives the opportunity to use analytical formula for the rate of ionization $\beta$ .

In Ref~\cite{Istomin:17} dense molecular plasma is studied, which arises
in the Earth's atmosphere at re-entrance of spacecrafts, for example.
On the other hand, the results of our work find application in the partially ionized
solar chromosphere and its sound wave heating.
Although these studies focus on different physical scenarios, high vs low dense plasma,
the common conclusion is however the same:
up to now the bulk viscosity has not been in the focus of plasma physics,
despite many interesting applications.
This is also noted in Ref.~\cite[Sec.~Introduction]{Istomin:17}:
\textit{However, in the majority of studies
(except Refs.~10 and 12--14) atomic species are considered
as structure-less particles. Under this assumption, no internal heat conductivity and bulk viscosity coefficients appear for atomic species, both neutral and ionized, which can cause the loss of accuracy in the transport terms evaluation.}

\section{Saha equation and plasma thermodynamics. 
Cold plasmas approximation}
\subsection{Ideal gas approximation. System of notations}

We analyze low density plasma for which its pressure is given by the ideal gas approximation
\begin{align}
p = n_\mathrm{tot}T, \quad
n_\mathrm{tot} = & \mathcal{N}_\mathrm{tot}n_\rho, 
\label{pressure}\\
& \mathcal{N}_\mathrm{tot } =  \mathcal{N}_a+\mathcal{N}_e,\quad
n_e=\mathcal{N}_e\,n_\rho,
\label{density}
\end{align}
where $\mathcal{N}_e$ is the number of electrons per hydrogen atom,
the electrons per unit volume is $n_e$ and $n_\mathrm{tot}$ is the total number of particles.
Introducing an atomic mass per hydrogen atom
\be
M^*=\sum_a \overline a_a M_a,
\ee
where $M_a$ is the atomic mass per atom $_a$ and
the mass density of the gas can be written as
\be
\rho=M^*n_\rho.
\ee
It is also convenient to introduce averaged mass of the plasma particles  including the electrons
\be
\langle M\rangle\equiv M^*/\mathcal{N}_\mathrm{tot}=\rho/n_\mathrm{tot},
\ee
and the thermal velocity per averaged plasma particles
\be
c_\mathrm{N}^2=\frac{T}{\langle M\rangle}
=\overline{p}/\overline{\rho}=\left(\frac{\partial p}{\partial \rho}\right)_{\!T}.
\ee
As a rule, with an overline we denote thermally averaged quantities,
meaning the system is in local thermodynamic equilibrium (LTE).
The notations follow the same ones from Ref.~\cite{ApJ:24}.

\subsection{The Saha equation}

Now we can analyze the Saha equation for equilibrium ionization~\cite{Saha:21}.
Each atom with an atomic number $Z_a$ can be in a different degree of ionization, and its probability to be in it is denoted with $r_{i,a}$, where $i=0,\dots ,Z_a$ shows the successive ionization state of the atom, i.e.
$i=0$ is for the neutral state, $i=1$ is for the first ionized state and so on.
In such a way, the volume number of the corresponding ions
\be
n_{i,a}=n_\rho\overline{a}_ar_{i,a},\qquad
\sum_{i=1}^{Z_a} \,r_{i,a}=1,
\label{normalization}
\ee
and the charge neutrality condition gives the electron density
\be
n_e=n_\rho\,\mathcal{N}_e,\qquad
\mathcal{N}_e=\sum_a \overline{a}_a \sum_{i=1}^{Z_a}
i\,r_{i,a}.
\label{ne_ai}
\ee
Often the corresponding concentration is used~\cite[Sec.~104]{LL5}
and it is straightforward to change the notations
\be
c_{i,a}\equiv \frac{n_{i,a}}{n_\mathrm{tot}},
\quad
c_{e}\equiv \frac{n_e}{n_\mathrm{tot}},
\quad
n_\mathrm{tot}=n_e+n_\rho\sum_a \overline a_a.
\ee
Within the framework of the up to now introduced notations, the Saha equation
for the equilibrium concentrations reads
in its standard form~\cite[Eq.~(104.2)]{LL5}
\be
\frac{\overline c_{i-1,a}}
{\overline c_{i,a}\,\overline c_e}
=\frac{\overline p}{T\tilde{n}_{i,a}},
\qquad
\overline p=\overline n_\mathrm{tot}
\overline T=
\overline{n}_{\mathrm{\rho}}
\overline{\mathcal{N}}_\mathrm{tot}
\overline{T},
\label{overline_p}
\ee
where following Ref.~\cite{ApJ:24}, we have introduced a chain of almost standard notations
\begin{align}
&
\tilde{n}_{i,a}\equiv\frac{g_{i,a}g_e}{g_{i-1,a}}n_{s,i,a},
\label{tilde_n_(1,a)}\\
&
n_{s,i,a}\equiv n_q\e^{-\iota_{i,a}},\quad
\iota_{i,a}\equiv\frac{I_{i,a}}{T},\quad
n_q\equiv\left(\frac{mT}{2\pi\hbar^2}\right)^{\!\!3/2}
\!\!,\nn
\end{align}
where $I_{i,a}$ is the $i$-th ionization potential of atom $a$,
and $g_{i,a}$ are the statistical weights of the corresponding ions.
Here we suppose that the plasma temperature is much higher than the fine energy splitting.
For hydrogen, for example, $I_{1,\mathrm{H}}=13.6\,\mathrm{eV}$,
$g_{1,\mathrm{H}}=2$, $g_{0,\mathrm{H}}=1$, 
and for the electrons we have two spin states and $g_e=2$.

It is numerically convenient to rewrite the Saha equation as
\be
\frac{\overline r_{i,a}}{\overline r_{i-1,a}}=\frac{\overline n_{i,a}}{\overline n_{i-1,a}}
=\frac{\tilde{n}_{i,a}}{\overline n_e},
\label{Saha_working}
\ee
and to express $\overline r_{i,a}$ by $\overline r_{i-1,a}$
starting with $r_{0,a}$ very small and $\overline{n}_e = \overline{n}_\rho/2$~\cite{ApJ:24}.
Then we can make the normalization \Eqref{normalization} and
after calculating all $\overline r_{i,a}$, we can again
calculate the electron concentration according to \Eqref{ne_ai}.
This procedure is repeated until convergence with a required accuracy is reached.
And last, we calculate the equilibrium pressure $\overline{p}$
using \Eqref{overline_p}.
Knowing the equilibrium degree of ionization,
in the next subsection we calculate all thermodynamic variables.

\subsection{Thermodynamic variables}

At known ionization probabilities,  we can easily calculate the ionization 
energy per atom for each atom
\begin{align}
E_a^{\mathrm{(ion)}}\equiv\sum_{i=1}^{Z_a} J_{i,a}r_{i,a},\qquad
J_{j,a}\equiv\sum_{i=1}^j I_{i,a}.
\end{align}
or in equilibrium
\be
\overline{E}_a^{\mathrm{(ion)}}\equiv\sum_{i=1}^{Z_a} J_{i,a}\overline{r}_{i,a}.
\ee
Then in order to calculate the enthalpy per unit mass $w$,
it is convenient to calculate the enthalpy per a hydrogen atom 
\be
M^* w=\mathcal{N}_e c_p T
+\sum_{a}\overline{a}_a
\left[
c_p T+\sum_{i} r_{i,a}J_{i,a},
\right], 
\label{general_w} 
\ee
where $c_p=5/2$ is the heat capacity per atom at constant pressure.
For the internal energy per hydrogen atom we have to substitute
$c_p$ with the heat capacity per constant volume per one atom $c_v=3/2$
\begin{align}
M^*\varepsilon
=&\, (\mathcal{N}_e+\mathcal{N}_\mathrm{atom}) c_vT
+\sum_a\overline{a}_a E_a^{\mathrm{(ion)}}
\label{energy_per_hydrogen_1}\\
=&\sum_a\overline{a}_a\!
\left[
\left(1\!+\!\sum_{i=1}^{Z_a}ir_{i,a}\right) \! c_v T
+\sum_{i=1}^{Z_a}r_{i,a}
\sum_{j=1}^{i}I_{j,a}\right].\nn
\end{align}
For brevity, in equilibrium overline $\overline{r}_{i,a}$ in all those formulae have been omitted.
In low frequency dynamics we can consider time dependent $r_{i,a}(t)$ and temperature $T(t)$, and as a consequence: time dependent internal energy 
$\varepsilon(t)$ and enthalpy $w(t)$ per unit mass.

Knowing these thermodynamic potentials,
we can easily calculate the heat capacity 
per unit mass from the derivatives~\cite[Sec.~16]{LL5}
\begin{align}
\mathcal{C}_p &\equiv 
\left(\dfrac{\partial w}{\partial T}\right)_{\!\! p},\\
\mathcal{C}_v &\equiv 
\left(\dfrac{\partial \varepsilon}{\partial T}\right)_{\!\! \rho}
=\left(\dfrac{\partial w}{\partial T}\right)_{\!\! \rho}
-\dfrac{1}{\rho}\left(\dfrac{\partial p}{\partial T}\right)_{\! \! \rho}.
\end{align}
Then the Jacobian~\cite[Eq.~(9)]{ApJ:24}
\be
 \mathcal{J} \equiv 
\frac{\partial (w,p)}{\partial (T,\rho)} 
= 
\left(\frac{\partial w}{\partial T}\right)_{\!\! \rho}
\left(\frac{\partial p}{\partial \rho}\right)_{\!\! T}
-\left(\frac{\partial w}{\partial \rho}\right)_{\!\! T}
\left(\frac{\partial p}{\partial T}\right)_{\!\! \rho}, 
\label{Jacobian} 
\ee
can also be calculated by numerical differences 
if the enthalpy is calculated with almost machine accuracy.

The sound velocity at evanescent frequency can be calculated by the standard
thermodynamic derivatives~\cite[Eq.~(A9)]{ApJ:24}
\begin{align}
c_0^2(T,\rho)&=
\left(\frac{\partial p}{\partial \rho}\right)_{\!\!s}
=\frac{\mathcal{J}}{\mathcal C_v} \\
& =\dfrac{\left(\dfrac{\partial w}{\partial T}\right)_{\!\! \rho}
\left(\dfrac{\partial p}{\partial \rho}\right)_{\!\! T}
-\left(\dfrac{\partial w}{\partial \rho}\right)_{\!\! T}
\left(\dfrac{\partial p}{\partial T}\right)_{\!\! \rho}}
{\left(\dfrac{\partial w}{\partial T}\right)_{\! \rho}
-\dfrac{1}{\rho}\left(\dfrac{\partial p}{\partial T}\right)_{\! \! \rho}}. \nn 
\end{align}
For our further analysis, this velocity is convenient to be represented 
by the generalized polytropic index
\begin{align}
\gamma_0&\equiv
\gamma_0(T,\rho)
\equiv
\dfrac{\left(\dfrac{\partial p}
{\partial \rho}\right)_{\!\!s}}
{\left(\dfrac{\partial p}{\partial \rho}\right)_{\!T}}
=\frac{c_0^2}{c_\mathrm{N}^2}
=\frac{\rho}{p}\left(\frac{\partial p}{\partial \rho}\right)_{\!\!s}
=\frac{\rho}{p} \cdot
\dfrac{\mathcal{J} }{\mathcal{C}_v}
\nn\\
& =\frac{\rho}{p}
\cdot
\dfrac{\left(\dfrac{\partial w}{\partial T}\right)_{\!\! \rho}
\left(\dfrac{\partial p}{\partial \rho}\right)_{\!\! T}
-\left(\dfrac{\partial w}{\partial \rho}\right)_{\!\! T}
\left(\dfrac{\partial p}{\partial T}\right)_{\!\! \rho}}
{\left(\dfrac{\partial w}{\partial T}\right)_{\! \rho}
-\dfrac{1}{\rho}\left(\dfrac{\partial p}{\partial T}\right)_{\! \! \rho}}. 
\label{gamma_thermodynamic_1}
\end{align}
In the formulae above, the overline denoting thermal equilibrium values have been omitted again.
The velocity $c_0$ is the sound velocity at low frequencies for which during an adiabatic compression of the plasma, the ionization degrees can follow the equilibrium Saha equation.

In the opposite limit, when the ionization-recombination processes are negligible
and the adiabatic compression is at constant chemical compound (isochema),
for the sound velocity we have the standard monoatomic value
\be
c_\infty^2=\gamma_\infty \, c_\mathrm{N}^2,\qquad 
\gamma_\infty=\gamma_a=5/3.
\ee 

Excluding these two limiting cases, the sound propagation with damping is
determined by the kinetics of the ionization-recombination processes.
For real frequencies, the wave amplitudes of the pressure and density are absorbed
by the imaginary part of the frequency dependent
wave number $k^{\prime\prime}(\omega)$
\begin{align}
&
\delta p=\Re(\hat{p}(x,t)),\qquad
\delta \hat{p}(x,t)\propto\exp(\im(\hat{k}x-\omega t)),\\
&
\delta \rho=\Re(\hat{\rho}(x,t)),\qquad
\delta \hat{\rho}(x,t)\propto\exp(\im(\hat{k}x-\omega t)),\\
&
\e^{\im\hat{k}x}=\e^{-k^{\prime\prime}x}\e^{\im k^{\prime}x},
\qquad \hat{k}=k^{\prime}+\im k^{\prime\prime}.
\end{align}

The central purpose of the present study is to calculate
the generalized polytropic index $\hat\gamma(\omega)$ describing the entropy production by the ionization-recombination processes parameterized by the complex phase velocity~\cite[Eq.~(81.8)]{LL6}
\begin{align}
\hat{c}_\mathrm{phase}^2 \equiv \frac{\delta \hat{p}}{\delta \hat{\rho}}=\frac{\omega^2}{\hat{k}^2}
=\hat{\gamma}(\omega)\,c_\mathrm{N}^2.
\end{align}
In order to analyze these processes for cold plasmas, in the next section we
shortly review the physics of the near threshold ionization cross-section
for the simplest possible reaction of the ionization of a hydrogen atom
\be
\rm e+H \rightarrow p+e+e.
\label{reaction}
\ee

\section{Wannier ridge resonances and Wannier 
near threshold ionization cross-section}

It is remarkable that the exact solution for the scattering in the Coulomb potential coincides with both Born and classical cross-section approximations.
The physics, however, is absolutely different.
At high energies we have slightly perturbed plane waves while the
classical approximation is applicable when the velocity $v$ of the scattering electron 
is much smaller than the Bohr velocity
\be
v_\mathrm{_B}\equiv\alpha_\mathrm{_S}\,c,
\qquad
\alpha_\mathrm{_S}\equiv\frac{e^2}{\hbar c},
\qquad
e^2\equiv\frac{q_e^2}{4\pi\varepsilon_0}.
\ee

On the other hand, the quasi-classical motion of two $s$-electrons is extremely interesting.
They move as mirror images with opposite radius vectors 
$\mathbf{r}_1=-\mathbf{r_2}$ and opposite momenta
$\mathbf{p}_1=-\mathbf{p_2}$ like two electrons in a Cooper pair.
This strongly correlated motion is an example of the inapplicability of 
the self-consistent Hartree-Fock approximation for this problem.
However, strong correlations lead to another simplification:
the effective Rydberg takes into account the mirror electron
\be
E_n=-\frac{\mathrm{R_{eff}}}{n^2},\qquad
\mathrm{R_{eff}}=\frac12 \left(Z-\frac14\right) mc^2\alpha_\mathrm{_S}^2,
\ee
where $Z$ is the charge of the rest ion and $m$ is the electron mass.
The velocity of light $c$ disappears
in all those formulae into the Sommerfeld fine splitting constant 
$\alpha_\mathrm{_S}$.
If we draw the effective potential of this 3-body system in some choice of variables, we observe a ridge. 
That is why those metastable resonances are called Wannier ridge resonances
and are observed by electron impact in noble gases~\cite{Wannier:53,Read:82}.

But how does an electron with small energy slightly above the threshold $I$ create the ionization reaction \Eqref{reaction}?
The two electrons escape quasi-classically with small energies and again
the mirror images are important.
This important problem was solved by Wannier in 1953~\cite{Wannier:53},
see also~\cite[Sec.~147 Near-threshold cross section behavior]{LL3},
who obtained for the cross-section of the reaction \Eqref{reaction}
\be
\sigma(\varepsilon) = \cw a_\mathrm{B}^2 
\left ( \frac{\varepsilon}{I} -1 \right )^{\!\! \mathrm{w}}.
\ee
The numerical values we use $\mathrm{w} \approx 1.18\sim 1$ and 
$C_\mathrm{W} \approx 2.7$ are obtained from the experimental study~\cite[Fig.~6]{McGowan:68}.
We consider as a relevant discussion for plasma physics to explain that the
Wannier formula is applicable within an acceptable accuracy for all
$s$-electrons ionized by electron impact. 
The energy of the classical motion is just the quantum ionization potential.

Performing the integration over equilibrium Maxwell distribution of electrons~\cite[Eq.~(5)]{PhysA}
\be
\beta = \langle v \sigma(\varepsilon) \rangle, 
\qquad  v = \sqrt{2\varepsilon/m},
\ee
we obtain for the ionization rate of this impact ionization  
\begin{align}
 \beta(T)
 &\approx \frac{2}{\sqrt{\pi}}C_\mathrm{W} \Gamma(\mathrm{w}+1)\frac{\e^{-\iota}}{\iota^{\mathrm{w}-1/2}}
 \beta_\mathrm{B},
 \label{beta_a}\\
\beta_\mathrm{B}&\equiv v_\mathrm{_B} a_\mathrm{_B}^2=6.126\times 10^{-15}\, \mathrm{m^3/s}.
 \nn
\end{align}
The use of the near-threshold cross-section is a good approximation only for cold plasmas with $T\ll I$ or $\iota \gg 1$.
In many physical problems pure hydrogen plasma allows an analytical solution and 
for the ionization rate of cold hydrogen plasma or which there is an acceptable \textit{ab initio} approximation already~\cite{PhysA}.
 
\section{Cold plasma kinetic equation and its solution for small harmonic density oscillations}

\subsection{General kinetic equation}

The cold plasma approximation $T\ll I_a$  or $\iota_{1,a} \gg 1$ can be easily checked by the solution of the Saha equation taking into account only the first ionized state of all elements in the cocktail $I_a \equiv I_{i=1,a}$ hence $\iota_a \equiv \iota_{i=1,a}$, and $\beta\rightarrow\beta_a$ in \Eqref{beta_a}.
We also have
\be
\sum_{i=2}^{Z_a}\overline{r}_{i,a}\ll1
\ee
and therefore it is convenient to re-denote
\be
\alpha_a\equiv r_{1,a}
\label{alpha1}
\ee
and use the approximation
\be
r_{0,a}\approx 1-\alpha_a, \quad r_{2,a}\approx 0,
\quad r_{3,a}\approx 0,\quad \dots \,. 
\label{alpha0}
\ee
For these relatively low temperatures we take into account only single electron
ionization 
\be
\mathrm{e}+a_{0} \longleftrightarrow 
a_1+ \mathrm{e +e}.
\ee
Our starting point is the general kinetic equation for the volume density of
the $i$-th ionized state of the ion of atom $a$,
$n_{i,a}(t)$
\begin{align}
\md_t n_{i,a} = n_{i,a} \frac{\md_t n_\rho}{n_\rho}
&+\beta_{i,a}(T)\left[n_en_{i-1,a}-\frac{n_{i,a}n_e^2}{\tilde{n}_{i,a}(T)}\right]
\label{kinetic_equation_i_a}\\
&-\beta_{i+1,a}(T)
\left[n_e n_{i,a}-\frac{n_{i+1,a}n_e^2}{\tilde{n}_{i+1,a}(T)} \right]. \nn
\end{align}
where $\beta_{i,a}$ is the rate of the single electron reaction
\be
\mathrm{a}^{i+}+\mathrm{e}\rightarrow\mathrm{a}^{(i+1)+}+2\mathrm{e}.
\ee
We suppose that the multi-ionization processes have negligible contribution
in this low temperature approximation.
Having taken the first ionized state, from \Eqref{kinetic_equation_i_a} we have
for $n_{1,a}(t)$
\begin{align}
\md_t n_{1,a} &= n_{1,a} \frac{\md_t n_\rho}{n_\rho}+
\beta_{a}(T)\left[n_en_{0,a}-\frac{n_{1,a}n_e^2}{\tilde{n}_{1,a}(T)}\right]
\label{kinetic_equation}
\end{align}
with $\beta_{a}(T)$ given by \Eqref{beta_a}, the
Saha density parameter $\tilde{n}_{1,a}(T)$ given by \Eqref{tilde_n_(1,a)}
and for $n_{0,a}(t)$
\be
\md_t n_{0,a} =-\md_t n_{1,a}
\ee
from \Eqref{alpha1} and \Eqref{alpha0}.
The temperature $T(t)$ and electron density 
\be
n_e(t)\approx\sum_{a}n_{1,a}(t)=
n_\rho(t)\sum_a\overline{a}_a\,r_{1,a}(t)
\label{ne_a1}
\ee
and are also time dependent.
The difference between \Eqref{ne_ai} and the expression above \Eqref{ne_a1}
is that now we take into account only single ionization
supposing that for cold plasmas higher ionization states are negligible. 

\subsection{Linearization for small density oscillations}
Our task is to study how small oscillations of the density
\begin{align}
&
\rho(t)=[1+\epsilon_\rho(t)]\,\overline\rho,\quad
\epsilon_\rho(t)=\Re(\hat{\epsilon}_\rho(t)),
\label{density_driver}\\
&
\hat\epsilon_\rho(t)\propto\e^{-\im\omega t},\qquad
|\hat\epsilon_\rho|\ll1.
\label{density_wave}
\end{align}
create small relative oscillations of all variables around their respective equilibrium values
\begin{align}
&
T=\Re(\hat{T}(t)),\quad\;\; \hat{T}(t)=(1+\hat{\epsilon}_{_T})\,\overline{T},\\
&
p=\Re(\hat{p}(t)),\qquad \hat{p}(t)=(1+\hat{\epsilon}_{p})\,\overline{p},\\
&
p=\Re(\hat{p}(t)),\qquad \hat{p}(t)=(1+\hat{\epsilon}_{p})\,\overline{p},\\
&
\alpha_a=\Re(\hat{\alpha}_a(t)),\quad \hat{\alpha}_a(t)=(1+\hat{\epsilon}_{a})\,\overline{\alpha}_a,\\
&
r_{i,a}=\Re(\hat{r}_{i,a}(t)),\;\;
\hat{r}_{i,a}(t)=(1+\hat{\epsilon}_{i,a})\overline{r}_{i,a}.
\end{align}
All these relative deviations $|\hat\epsilon|\ll1$ are small and 
$\propto\e^{-\im\omega t}$.
An alternative system of notations is to introduce the deviations
\begin{align}
&
\hat{T}=\overline T+\delta\hat{T},\qquad 
\hat{\epsilon}_{_T}=\delta\hat{T}/\overline T,\\
&
\hat{p}=\overline p+\delta\hat{p},\qquad \;\;
\hat{\epsilon}_{p}=\delta\hat{p}/\overline p,\\
&
\hat{\rho}=\overline \rho+\delta\hat{\rho},\qquad \;\;
\hat{\epsilon}_{\rho}=\delta\hat{\rho}/\overline \rho.
\end{align}
It is convenient to work with linearized equations as complex numbers can be used
and their real parts $\Re$ to be taken after the calculation.

In order to derive the linearized kinetic equation,
 we have to substitute these small deviations into the 
general kinetic equation \Eqref{kinetic_equation_i_a}
and finally in its cold plasma approximation \Eqref{kinetic_equation}.
Omitting the complete derivation, we now mention only some technical details.

The obtained relations between the variations of the probabilities in the cold plasma approximation are
\begin{align}
&
r_{1,a}=\alpha_{a},\qquad r_{0,a}= 1-\alpha_a, \qquad 
r_{0,a}+r_{1,a}=1,\\
&
r_{1,a}=\overline r_{1,a} (1+\epsilon_{a}), \qquad
r_{0,a}=\overline r_0 (1+\epsilon_{0,a}),\\
&
\epsilon_{0,a}=-\frac{\overline r_{1,a}}{\overline r_{0,a}}\epsilon_a
=-\frac{\overline\alpha_a}{1-\overline\alpha_a}\epsilon_a,\quad
\epsilon_a\equiv\epsilon_{1,a}= \Re{(\hat \epsilon_a)}.
\end{align}

Now we have to describe a chain of auxiliary notations:
1) Let us introduce 
\be
\overline{X}_a=\beta_a(\overline{T})\,\overline{n}_{0,a}\,\overline{n}_e
\ee
and take into account that
\be
\overline{n}_{0,a}=\overline{n}_\rho\,\overline{a}_a\,(1-\overline\alpha_a),
\quad
\overline{n}_\rho=\overline{\rho}/M^*,
\quad
\overline{n}_e=\overline{n}_\rho\,\overline{\mathcal{N}}_e.
\ee
2) As we suppose that the density fluctuations are sinusoidal \Eqref{density_wave},
for the time derivatives, following Heaviside, we introduce the imaginary variable 
\be
\mathrm{D}=-\im\omega,\quad \md_t\hat{n}_{1,a}=\mathrm{D}\,\hat{n}_{1,a},
\quad \md_t\hat{n}_{\rho}=\mathrm{D}\,\hat{n}_{\rho},\quad \dots \, .
\ee
3) For the oscillations of the Saha densities $\tilde{n}_{1,a}(T(t))$  we have to take in the beginning that
\be
\delta\hat{T}^\prime \,\frac{\md\tilde n_{1,a}}{\md T}
=(c_v+\iota_{1,a})\,\tilde n_{1,a}\hat{\epsilon}_{_T},\qquad
\hat{\epsilon}_{_T}\equiv\frac{\delta\hat{T}}{\overline{T}}.
\ee
4) For the calculations of the temperature oscillations $\delta\hat{T}$ we have to take into account the energy conservation.
Let $\varepsilon$ be the internal energy per unit mass, then
we take the general expression for all possible ionization degrees
\Eqref{energy_per_hydrogen_1}.

For the general case, the set of the kinetic equations for $n_{i,a}$ is completed by the 
equation for the temperature variation $\delta\hat{T}^\prime$ derived from the energy conservation of the liquid particle with unit mass
\be
\md \varepsilon=-p\,\md \mathcal{V},\qquad
\mathcal{V} \equiv 1/\rho.
\ee
Dividing $\md \varepsilon$ 
by $\md t$ gives the differential equation
\be
\dot{\varepsilon}=-p\dot{\mathcal{V}}
=\frac{p}{\rho^2}\,\dot{\rho}
=\frac{p}{\rho}\,
\md_t\epsilon_\rho,\qquad
\md_t=\frac{\md}{\md t},
\ee
which can be written as
\begin{align}
&\frac1{M^*}
\left[
\mathcal{N}_\mathrm{tot}c_v\dot{T}
+c_vT\dot{\mathcal{N}}_e
+T\sum_{a}\overline{a}_a \sum_{i=1}^{Z_a}
\overline{r}_{i,a}\dot{\epsilon}_{i,a}
l_{i,a}
\right],
\nn\\
&
=\frac{T}{M^*}\mathcal{N}_\mathrm{tot}\,
\md_t\epsilon_\rho, 
\qquad
l_{i,a}\equiv J_{i,a}/T=\sum_{j=1}^i\iota_{i,a}.
\label{temporary_equation}
\end{align}
This linearized equation between the first time derivatives 
gives the final link between the small relative variations of the temperature, density and ionization degrees.
Re-denoting $\md_t$ with over-dots
\begin{align}
\epsilon_{_T}=\frac{\epsilon_\rho}{c_v}
-\frac{\overline{\mathcal{N}}_e}
{\overline{\mathcal{N}}_\mathrm{tot}}\epsilon_e
-\frac{1}{c_v\overline{\mathcal{N}}_\mathrm{tot}}
\sum_a \overline{a}_a \sum_{i=1}^{Z_a}
\overline{r}_{i,a}l_{i,a}\epsilon_{i,a},
\label{complex_iota_rho_1}
\end{align}
where the overline denoting thermodynamic 
equilibrium averaging can be omitted in the further linearized equations.

In such a way, from the general kinetic equation
\Eqref{kinetic_equation_i_a}, we have derived the linearized
equation for the relative deviations from the equilibrium
\begin{align}
&
\hat \epsilon_{i,a}
=\frac{\overline{n}_e}{\mathsf{D}}\left\{
-\left(\beta_{i,a}\,
\frac{
\overline{r}_{i-1,a}}{\overline{r}_{i,a}}\right)
(\hat \epsilon_{i,a} + \hat \epsilon_e - \hat \epsilon_{i-1,a})
\right.
\label{Boltzmann_1}\\
&
\qquad\qquad
+\beta_{i+1,a}\,(\hat \epsilon_{i+1,a} + \hat \epsilon_e - \hat \epsilon_{i,a})
\nn \\
&
\left.
+\left[
(c_v+\iota_{i,a})\left(\beta_{i,a}\,\frac{
\overline{r}_{i-1,a}}{\overline{r}_{i,a}}\right)
-(c_v+\iota_{i+1,a})\,\beta_{i+1,a}
\right] \hat \epsilon_{_T}
\right\}.
\nn
\end{align}

Additionally the charge neutrality condition of the plasma gives for the electron density
\begin{align}
\mathcal{N}_e & =\frac{n_e}{n_\rho}
=\sum_a\overline{a}\sum_{i=1}^{Z_a}
i\overline{r}_{i,a}(1+\epsilon_{i,a}),
\nn
\\
\dot{\mathcal{N}}_e & =\frac{n_e}{n_\rho}
=\sum_a\overline{a}\sum_{i=1}^{Z_a}
i\overline{r}_{i,a}\dot{\epsilon}_{i,a},
\nn \\
\mathcal{N}_e^\prime &
=\int \dot{\mathcal{N}}_e\,\md t
=\mathcal{N}_e-\overline{\mathcal{N}}_e,
\nn \\
& =\sum_a\overline{a}\sum_{i=1}^{Z_a}
i\overline{r}_{i,a}\epsilon_{i,a}
=\overline{\mathcal{N}}_e\,\epsilon_e,\nn\\
\hat\epsilon_e & =\frac{\hat{\mathcal{N}}_e^\prime}
{\overline{\mathcal{N}}_e}
=\frac{\sum_a\overline{a}\sum_{i=1}^{Z_a}
i\overline{r}_{i,a}\hat\epsilon_{i,a}}{\sum_a\overline{a}\sum_{i=1}^{Z_a}
i\overline{r}_{i,a}}.
\label{electron_density_1}
\end{align}
One can also introduce relaxation rates having dimension of frequency
\be
f_{i,a}\equiv\frac1{\tau_{i,a}}\equiv\beta_{i,a}\overline{n}_e,\qquad
i \in [1,\,Z_a].
\label{relax}
\ee
For completeness, we have to introduce `boundary conditions'
$f_{-1,a}=f_{Z_a+1,a}=0$.
Considering the physical meaning
\be
\tau_{i,a}^{(1/2)}
\equiv\ln(2)\,\tau_{i,a}
\ee
is the semi-decay period
one ion a$^{(i-1)+}$ to be ionized by an electron impact and to become a$^{i+}$.

One can trace that $\hat{\epsilon}_{_T}$ in \Eqref{Boltzmann_1} 
arises from the time dependent temperature in the Saha densities 
$\tilde{n}_{i,a}$ from \Eqref{kinetic_equation_i_a}.
The substitution of $\hat{\epsilon}_{_T}$ from the equation of the 
temperature oscillations \Eqref{complex_iota_rho_1}
into the linearized kinetic equation 
\Eqref{Boltzmann_1} for the relative oscillations of the concentrations gives
\Eqref{electron_density_1}  for
$\hat\epsilon_e$  and in this way we have obtained the
complete set of equations only for $\hat{\epsilon}_{i,a}$, i.e. we have eliminated the relative temperature variation.

\subsection{Numerical method for singly ionized cold plasmas}

The general procedure for derivation of a set of linear equations for
$\hat\epsilon_{i,a}$ we give in explicit form only in the cold 
plasma approximation which we now study numerically.
It is convenient to introduce the following simplified notations
\begin{align}
&
\hat\epsilon_a\equiv \hat{\epsilon}_{1,a}, 
\quad \hat \epsilon_{0,a} = - \frac{\overline{\alpha}_a} {1-\overline{\alpha}_a}
\hat \epsilon_a, \\
&
\hat \epsilon_{a,2}=\hat \epsilon_{a,3}=\dots=\hat \epsilon_{a,i\,>\,1}=0,\\
&
f_{a}\equiv\frac1{\tau_{a}}\equiv\beta_{1,a}\overline{n}_e.
\end{align}
If we treat only ground and 1-st ionized state ($i \in [0, 1]$),
\Eqref{ne_ai} gives
\begin{align}
\mathcal{N}_e & =\sum_a \overline{a}_a \sum_{i=1}^{Z_a} i \,r_{i,a} \approx
\sum_a \overline{a}_a \, (1 \, \,r_{1,a})=
\sum_a \overline{a}_a  \alpha_a \nn \\
& = \sum_a \overline{a}_a \overline{\alpha}_a (1+\epsilon_{1,a}) =
\overline{\mathcal{N}}_e (\epsilon_e+1), \nn \\
\mathcal{\overline{N}}_e & = \sum_a \overline{a}_a  \overline{\alpha}_a.
\label{Ne_alpha}
\end{align}
This of course is valid in the case for low temperature plasma where $\iota_{1,a} \equiv I_{1,a}/T \gg 1$.
The relation for the charge neutrality \Eqref{electron_density_1}
now gives
\be
\hat \epsilon_e \approx \langle \hat \epsilon_a \rangle_a =
\dfrac{\sum_a \overline{a}_a \overline{\alpha}_a \hat \epsilon_{a}}
{\sum_a \overline{a}_a \overline{\alpha}_a}.
\label{epsilon_e_1}
\ee
Analogously, the general expression for the temperature oscillation \Eqref{complex_iota_rho_1} takes the form
\be
\hat \epsilon_{_T}=\frac{\epsilon_\rho}{c_v}
-\frac{\overline{\mathcal{N}}_e}
{\overline{\mathcal{N}}_\mathrm{tot}} \hat \epsilon_e
-\frac{1}{c_v\overline{\mathcal{N}}_\mathrm{tot}}
\sum_a \overline{a}_a \overline{\alpha}_a \iota_a \hat \epsilon_a.
\label{epsilon_T}
\ee
Substituting here $\epsilon_e$ from \Eqref{epsilon_e_1} we obtain
\be
\hat \epsilon_{_T} = \frac{\epsilon_\rho}{c_v} 
-\frac{1}{c_v \overline{\mathcal{N}}_\mathrm{tot}}
\sum_a \overline{a}_a \overline{\alpha}_a (c_v+\iota_a) \hat \epsilon_a.
\label{epsilon_T_2}
\ee
Next, we substitute this relative temperature oscillation 
$\epsilon_{_T}$ and the electron density oscillation 
$\hat \epsilon_e$ from \Eqref{epsilon_e_1} into the general kinetic equation 
\Eqref{Boltzmann_1} which for $\hat\epsilon_{1,a} \equiv \hat\epsilon_a$ reads 
\begin{widetext}
\be
\hat{\epsilon}_a=\dfrac{1-\overline\alpha_a}
{1-\im\omega\tau_a\overline\alpha_a}
\left\{\dfrac{\iota_a}{c_v}\hat\epsilon_\rho 
-\dfrac1{\overline{\mathcal{N}}_\mathrm{tot}}
\left[
\left(1+\dfrac{\iota_a}{c_v}\right)
\sum_{b=1}^{a_\mathrm{max}} \iota_b\overline{a}_b\overline\alpha_b\hat\epsilon_b
+ \left(\iota_a +c_v+\dfrac1{\overline{c}_e}
\right)
\sum_{b=1}^{a_\mathrm{max}} \overline{a}_b\overline\alpha_b\hat\epsilon_b
\right]\right\}. 
\label{Bash_Equation_1}
\ee
\end{widetext}
Here for each $a=$ H, He, C, Mg, Si, Fe we have to solve the corresponding 
linear equation for $a_\mathrm{max} = 6$ variables $\hat{\epsilon}_a$.
In order to alleviate the numerical programming, we introduce some convenient notations
\begin{align}
&
A_a=\overline{a}_a\overline{\alpha}_a,
\qquad  B_a=\iota_a\overline{a}_a\overline{\alpha}_a
=\iota_aA_a,\\
& 
\hat \Gamma_a=\dfrac{1-\overline\alpha_a}{1-\im\overline\alpha_a\tau_a},
\qquad a=1,\,\
2,\,\dots a_\mathrm{max},\\
&
C_a=\left(\iota_a+c_v+1/\overline{c}_e\right)/\overline{\mathcal{N}}_\mathrm{tot},\\
&
D_a=\left(1+\iota_a/c_v\right)/\overline{\mathcal{N}}_\mathrm{tot},
\end{align}
the row matrices
\begin{align}
&
\sf{L}_A=
\begin{pmatrix}
A_1&A_2&\dots&A_{a_\mathrm{max}}\\
\end{pmatrix},\\
&
\sf{L}_B=
\begin{pmatrix}
B_1&B_2&\dots&B_{a_\mathrm{max}}
\end{pmatrix},
\end{align}
the square matrices $a_\mathrm{max}\times a_\mathrm{max}$
\begin{align}
&
\hat{\sf{M}}_A=
\begin{pmatrix}
\hat \Gamma_1C_1\sf{L}_A\\
\hat \Gamma_2C_2\sf{L}_A\\
\dots\\
\hat \Gamma_{a_\mathrm{max}}C_{a_\mathrm{max}}\sf{L}_A\\
\end{pmatrix}
,\quad
\hat{\sf{M}}_B=
\begin{pmatrix}
\hat \Gamma_1D_1\sf{L}_B\\
\hat \Gamma_2D_2\sf{L}_B\\
\dots\\
D_{a_\mathrm{max}}\sf{L}_B\\
\end{pmatrix},\\
& \hat{\sf{M}}=\openone+\hat{\sf{M}}_A+\hat{\sf{M}}_B,
\end{align}
and the column matrices
\be
\hat{\sf{R}}=
\begin{pmatrix}
\hat \Gamma_1\iota_1\\
\hat \Gamma_2\iota_2\\
\dots\\
\hat \Gamma_{a_\mathrm{max}}\iota_{a_\mathrm{max}}\\
\end{pmatrix}
\frac{\epsilon_\rho}{c_v}
,\qquad
\hat\epsilon=
\begin{pmatrix}
\hat\epsilon_1\\
\hat\epsilon_2\\
\dots\\
\hat\epsilon_{a_\mathrm{max}}\\
\end{pmatrix}.
\ee
Now the solution of \Eqref{Bash_Equation_1} reads
\be
\hat{\sf{M}}\hat\epsilon=\hat{\sf{R}},\qquad
\hat\epsilon=\hat{\sf{M}}^{-1} \hat{\sf{R}},
\label{epsM_1}
\ee
and \Eqref{epsilon_e_1} and \Eqref{epsilon_T_2} can be correspondingly rewritten as
\begin{align}
&
\hat\epsilon_e=\sf{L}_A\cdot\hat\epsilon/\overline{\mathcal N}_e,\\
&
\hat\epsilon_{_T}=\hat\epsilon_\rho/c_v
+\overline c_e\hat\epsilon_e
+\sf{L}_B\cdot\hat\epsilon/\overline{\mathcal{N}}_\mathrm{tot},
\end{align}
where the central dot ``$\cdot$'' denotes matrix multiplication.

After the numerical solution of the kinetic equation,
we calculate all physical variables.
The density deviation from equilibrium \Eqref{density}
\be
n_\mathrm{tot} = \overline{n}_\mathrm{tot} (1+\epsilon_n) = n_\rho \mathcal{N}_\mathrm{tot}
\ee
is easily derived in terms of the electron and mass density deviations
\be
\epsilon_n = \epsilon_\rho+\overline{c}_e\,\epsilon_e,
\ee
where the ratio 
$\overline{c}_e\equiv\overline{\mathcal{N}}_e/\overline{\mathcal{N}}_\mathrm{tot}\in [0,\,1]$
is the electron concentration in equilibrium.
Analogously for the pressure \Eqref{pressure}
\be
p=\overline{p}\,(1+\epsilon_p) =
n_\mathrm{tot}\,T = \overline{n}_\mathrm{tot}\,(1+\epsilon_n)\,T_0 \,(1+\epsilon_{_T})
\ee
and its deviation from equilibrium
\begin{align}
\hat \epsilon_p & = \hat \epsilon_n + \hat \epsilon_{_T} =
\epsilon_\rho + \hat \epsilon_{_T} + \overline{c}_e \hat \epsilon_e,\\
\rho & = M^* n_\rho = \overline{\rho}\,(1+\epsilon_\rho), \qquad
M^* \equiv \sum_a \overline{a}_a M_a.
\end{align}
After this kinetic analysis, we can address 
the initial hydrodynamic problem
of sound propagation in homogeneous cold plasmas.

\subsection{Generalized polytropic index for cold plasmas}

Having calculated all necessary deviations from equilibrium, 
we can further proceed our analysis with the sound waves absorption.
For that purpose following~\cite[Sec.~81, Second viscosity]{LL6},
at real frequency $\omega$ we can introduce a
complex wave vector $\hat{k}=k^\prime+\im k^{\prime\prime}$
and a complex phase velocity
\begin{align}
\hat{c}^2_\mathrm{phase}& \equiv \frac{\omega^2}{\hat{k}^2(\omega)} 
\equiv\left(\frac{\partial p}{\partial \rho}\right)_{\!\!\omega }
\equiv\frac{\delta\hat{p}}{\delta\hat{\rho}}
= c_\mathrm{_N}^2\, \hat{\gamma} (\omega), \\
\hat{\gamma}(\omega) & \equiv \frac{\hat{\epsilon}_p}{\hat{\epsilon}_\rho}
 = 1 + \frac{\hat{\epsilon}_{_T} + \overline{c}_e \hat{\epsilon}_e}{\hat{\epsilon}_\rho} \\
& =  \gamma_a -\frac{1}{c_v\overline{\mathcal{N}}_\mathrm{tot} \hat{\epsilon}_\rho}
\sum_{a=1}^{a_\mathrm{max}} \overline{a}_a \overline{\alpha}_a 
\iota_a \hat{\epsilon}_a
\label{general_gamma}\\
&=  \gamma_a -\frac{\sf{L}_B\cdot\hat\epsilon}
{c_v\overline{\mathcal{N}}_\mathrm{tot} \hat{\epsilon}_\rho}
=\gamma^{\prime}+\im\gamma^{\prime\prime},
\label{complex_gamma_1}
\end{align}
where $c_\mathrm{_N}^2=\overline p/\overline\rho.$

As a test for the working programs we have to check that for very low
and very high frequencies the polytropic index is real only.
For high frequencies it reaches the atomic value $\gamma_a$ corresponding
to negligible chemical reactions.
For low frequencies we have a sophisticated thermodynamic expression
\Eqref{gamma_thermodynamic_1} \cite[Eq.~(8)]{ApJ:24}
\begin{align}
& \hat{\gamma}(\omega \rightarrow \infty) = \gamma_a \equiv 5/3, \qquad
\gamma^{\prime\prime}(\infty) \rightarrow 0, \quad
\gamma^{\prime}(\infty) \rightarrow \gamma_a, \nn \\
& \hat{\gamma}(\omega \rightarrow 0) \approx  \gamma^\prime(0)=\gamma_0, 
\qquad
\gamma^{\prime\prime}(0) \rightarrow 0.
\end{align}
For the linear analysis in the programs we can choose $\epsilon_\rho=1$.
For some purposes it is instructive also to introduce relative ``polytropic'' index
\be
\hat\gamma_\mathrm{rel}\equiv\dfrac{\hat\gamma(\omega)}{\gamma_\infty}
=1-\frac{\sf{L}_B\cdot\hat\epsilon}
{c_p\,\overline{\mathcal{N}}_\mathrm{tot} \, \hat{\epsilon}_\rho},
\label{gamma_rel_1}
\ee
and for the complex ``phase'' velocity we now have
\begin{align}
\hat{c}^2_\mathrm{phase} =
\frac{\omega^2}{\hat{k}^2(\omega)}=\hat\gamma_\mathrm{rel}(\omega)\,c_\infty^2.
\end{align}
Having an explicit expression for $\hat\gamma_\mathrm{rel}(\omega)$,
we can calculate the complex wave-vector
\be
\hat k=\frac{\omega}{c_\infty\sqrt{\hat\gamma_\mathrm{rel}(\omega)}}
=\frac{\omega}{c_\mathrm{N}\sqrt{\hat\gamma(\omega)}}
=k^\prime+\im k^{\prime\prime}
\label{hat_k}
\ee
and its imaginary part gives the sound wave damping 
created by the bulk viscosity.

\section{Kinetic equation results analysis}

\subsection{Mandelstam-Leontovich and Cole-Cole approximations}

\begin{figure}[ht]
\centering
\includegraphics[scale=0.5]{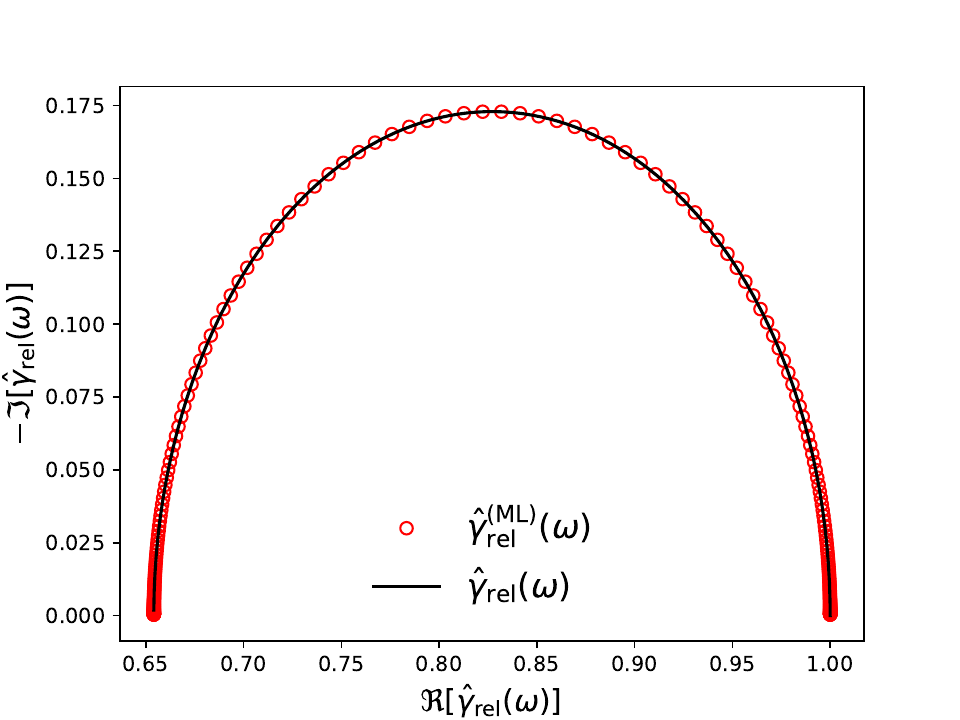}
\caption{Semicircular Cole-Cole~\cite{ColeCole} plot, or
Mandelstam-Leontovich (ML) fit of the relative polytropic index 
$\hat\gamma_\mathrm{rel}=\hat\gamma/\gamma_a$
for homogeneous plasma according to the AL08 profile~\cite{Avrett:08} and
the 6 component cocktail consisting of H, He, C, Mg, Si, Fe.
On the abscissa the real part $\Re(\hat\gamma_\mathrm{rel})$  is given
while in the ordinate it is the imaginary part
$-\Im(\hat\gamma_\mathrm{rel})$.
The line shows the exact calculation according to \Eqref{gamma_rel_1}, while
the open circles $\bigcirc$ correspond to the ML fit according to \Eqref{M-L_1}
with fitting parameter parameter $\Gamma_{0,\infty}=0.654$.
This arch is analogous to the Cole-Cole plot in the physics of liquids and dielectrics~\cite{ColeCole}.
}  
\label{fig:gama}
\end{figure}

We have solved the low temperature kinetic equation
\Eqref{epsM_1} for the solar plasma and calculated
the dimensionless ratio
\be
\Gamma_{0,\infty}\equiv\frac{\gamma_0}{\gamma_\infty}
=\dfrac{c_0^2}{c_{\infty}^2}=0.654<1.
\ee
The calculated frequency dependence 
$\hat\gamma(\omega)=\gamma^\prime+\im\gamma^{\prime\prime}$
deserves to be compared with 
the Mandelstam-Leontovich (ML) single time-constant 
approximation cf.~\cite[Eq.~81.8]{LL6}
\begin{align}
&
\hat\gamma_\mathrm{rel}^{\mathrm{(ML)}}(\omega)
=\dfrac{\Gamma_{0,\infty}-\im\omega\tau}{1-\im\omega\tau}
=\dfrac{\Gamma_{0,\infty}+\tilde{s}\tau}{1+\tilde{s}\tau},
\label{M-L_1}
\end{align}
where $\tilde{s}\equiv\im\omega$.
The comparison is represented in the Argand plot 
$\gamma^\prime$ versus $\gamma^{\prime\prime}$ drawn in 
\Fref{fig:gama}.
We are surprised that the results practically coincide.
The parameter $\Gamma_{0,\infty}$ 
related with the parameter $\mathcal{M}_\mathrm{RS} $ used 
by ~\citet{Rudenko:77} can be also expressed as
\be
\Gamma_{0,\infty}\equiv\ \frac{1}{1+\mathcal{M}_\mathrm{RS}}, \qquad
\mathcal{M}_\mathrm{RS} \equiv \frac{c_\infty^2}{c_0^2} -1,
\ee
see also Ref.~\cite[Eq.~(99)]{PhysA}.
The excellent accuracy of the ML fit 
with fitting parameter time-constant $\tau$
gives a simple interpretation
of the both low- and high-frequency limits:
we have a simple Pad\'e approximation.
We can consider the oscillation of the density as a perturbation
and the pressure as a linear response.
According to the causality principle, the generalized susceptibility 
$\hat p/\hat \rho\propto\hat\gamma(\omega)$
must be analytical in the upper semi-plane of the complex frequency
$\Im(\omega)>0$ and just along the imaginary semi-axis $\omega=\im \tilde s$
with $\tilde s>0$, the compressibility must be real and positive.
In the lower imaginary semi-plane 
$\hat\gamma_\mathrm{rel}^\mathrm{{ML}}(\omega)$
has 
a zero at $\omega=-\im/\tau$ and
a pole at $\omega=-\im\Gamma_{0,\infty}/\tau$
in agreement with the general causality principle 
\cite[Sec.~123 Generalized susceptibility]{LL5}.

\begin{figure}[ht]
\centering
\includegraphics[scale=0.5]{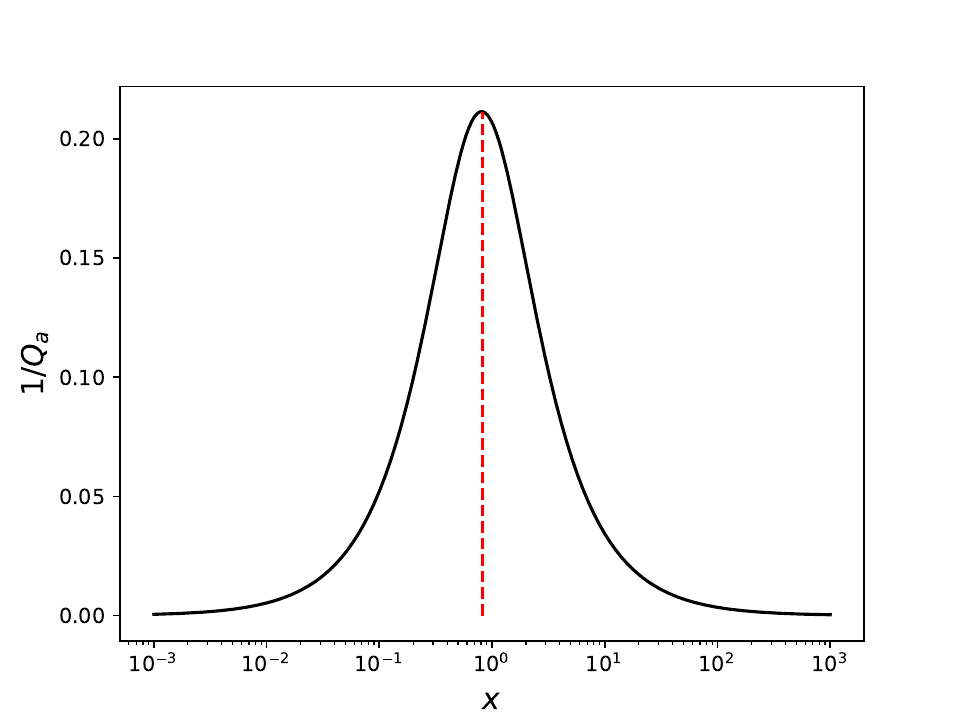}
\caption{
Reciprocal Q-factor of the ML approximation as a function
of $x \equiv \omega\tau$ according to \Eqref{Q_fit_1} and \Eqref{M-L_1}
$1/Q_{a,\mathrm{max}} \equiv Q_{a,\mathrm{min}} = 0.21$
at $x_\mathrm{max}=0.8235$.
Independently using \Eqref{Q_factor_kinetics_1}, we obtain 
$\omega_\mathrm{max}=0.9127$~\textmu rad/s 
and the corresponding time-constant is determined as
$\tau_\mathrm{max}=x_\mathrm{max}/\omega_\mathrm{max} \approx 10.5$~days.
In the general case
$1/\mathcal{Q}^{\mathrm{(ML)}}$ 
has a maximum at 
$x_\mathrm{max}=\omega_\mathrm{max} \tau_\mathrm{max}=\sqrt{\Gamma_{0,\infty}}=c_0/c_\infty$,
cf. \cite[Sec.~81]{LL6}.
Here we have plotted only the $1/\mathcal{Q}^\mathrm{(ML)}_a(\omega)$ from \Eqref{Q_fit_1} since the same calculation from \Eqref{Q_factor_kinetics_1} coincides with it as the Cole-Cole plot in \Fref{fig:gama}
}  
\label{fig:recQ}
\end{figure}

One can introduce also Q-factor for the sound waves
damped by the bulk viscosity
\be
Q_a(\omega)\equiv\frac{k^\prime}{2k^{\prime\prime}}
=\frac{\Re\left(1/\sqrt{\hat\gamma_\mathrm{rel}(\omega)}\right)}
{\Im\left(2/\sqrt{\hat\gamma_\mathrm{rel}(\omega)}\right)}\gg1.
\label{Q_factor_kinetics_1}
\ee
This factor is always large
$Q_a=k^\prime/2k^{\prime\prime}\gg1$.
One can define the real sound velocity~\cite[Sec.~81, Second viscosity]{LL6}
\be
c_\mathrm{s}(\omega)\equiv c_\infty\Re\left(\sqrt{\hat{\gamma}_\mathrm{rel}}\right)
=c_\mathrm{_N}\Re\left(\sqrt{\hat{\gamma}}\right)
\approx\Re(\hat{c}_\mathrm{phase}).
\ee
For the large enough values of the Q-factor,
one can represent relative polytropic index
\begin{align}
&
\hat\gamma_\mathrm{rel}(\omega)
=\gamma_\mathrm{rel}^\prime+\im\gamma_\mathrm{rel}^{\prime\prime}
=|\hat\gamma_\mathrm{rel}|\e^{\im\varphi_\gamma}
\approx\gamma_\mathrm{rel}^\prime(1+\im\varphi_\gamma),\\
&
\hat k=k^\prime+\im k^{\prime\prime}
=|k|\e^{\im\varphi_k}\approx
(1+\im/2Q_a)k^\prime,\\
&
|\varphi_\gamma|\ll1,\qquad\varphi_k=-\frac{1}{2}\varphi_\gamma
=\frac{1}{2Q_a}.
\end{align}
The general formula for the wave-vector
frequency dependence of the complex wave-vector, 
cf. Ref.~\cite[Eq.~(81.9-10)]{LL6}
\begin{align}
\hat k \equiv & \, k^\prime+\im k^{\prime\prime} \approx \frac{\omega}{c_\mathrm{s}}
+\im\frac{\omega^2\zeta^{\prime}(\omega)}{2\rho_0c_\mathrm{s}^3}
\label{complex_k}\\
=&\left(1+\im\frac{\omega\zeta^\prime}{2\rho c_\mathrm{s}^2}\right)
\frac{\omega}{c_\mathrm{s}}=\left(1-\im\frac{\varphi_\gamma}{2}\right)k^\prime
\label{semi_final}
\end{align}
gives the low frequency limit \cite[Eq.~(81.9)]{LL6}
\be
\hat k(\omega\ll1/\tau)\approx\frac{\omega}{c_0}
\left[1+\im \omega\tau\left(\frac{1}{\Gamma_{0,\infty}}-1\right)
\right]
\ee
and correspondingly the high-frequency limit
\cite[Eq.~(81.10)]{LL6} 
\be
\hat k(\omega\gg1/\tau)
=k^\prime+\im k^{\prime\prime}
\approx\frac{\omega}{c_\infty}
\left[1+\im \left(1-\Gamma_{0,\infty}\right)
\right]
\ee
for the ML approximation \Eqref{M-L_1}.
The frequency dependent reciprocal Q-factor of the ML approximation
\be
\frac1{Q_a^\mathrm{(ML)}(\omega)}
\equiv
-\frac
{2\Im\left(\sqrt{\hat\gamma_\mathrm{rel}^\mathrm{(ML)}(\omega)}\right)}
{\Re\left(\sqrt{\hat\gamma_\mathrm{rel}^\mathrm{(ML)}(\omega)}\right)}
\label{Q_fit_1}
\ee
is depicted in \Fref{fig:recQ}.
This function $\left[ Q_a^\mathrm{(ML)}\right]^{-1}$
has maximum at the argument
$x\equiv\omega\tau=\sqrt{\Gamma_{0,\infty}},$
where the parameter 
$\Gamma_{0,\infty}=c_0^2/c_\infty^2=0.654$
is determined by the ML fit represented in
\Fref{fig:gama}.

Let us describe a possible procedure of determination of the time-constant 
$\tau_\mathrm{max}$.
First we calculate 
$\gamma_0=\Re(\hat\gamma(\omega\rightarrow 0))$
according to \Eqref{complex_gamma_1}.
For high enough frequencies we have an important test
$\gamma_\infty=\Re(\hat\gamma(\omega\rightarrow\infty)) \equiv \gamma_a = 5/3$,
while for both very low and very high frequencies the imaginary part of
$\hat\gamma$ is negligible and hence at both limits
$\hat \gamma = \Re(\hat\gamma)$
and $\Im(\hat\gamma)=0$.
Then we calculate the ratio 
$\Gamma_{0,\infty}=\gamma_0/\gamma_\infty=0.654$.
With this parameter we plot the reciprocal quality factor 
$1/Q_a^\mathrm{(ML)}$
according to
\Eqref{Q_fit_1} where $\hat\gamma^{\mathrm{(ML)}}(x \equiv \omega\tau)$
is taken from \Eqref{M-L_1}.
This function has a maximum at $x_\mathrm{max}=\sqrt{\Gamma_{0,\infty}}$ shown in \Fref{fig:recQ}.
Independently, we calculate the reciprocal Q-factor $1/Q_a(\omega)$ 
from the solution of the kinetic equation \Eqref{Q_factor_kinetics_1}.
Next, we determine at which frequency $1/Q_a(\omega_\mathrm{max})$
is maximal and the time-constant 
\be 
\tau_\mathrm{max}=x_\mathrm{max}/\omega_\mathrm{max}
\label{tau_max_1}
\ee 
is calculated.
The exact value of $\tau_\mathrm{max}$ is irrelevant for the Argand plot in \Fref{fig:gama} where the exact complex function
$\hat\gamma$ and its ML approximation
$\hat\gamma^{\mathrm{(ML)}}$ are represented.

In the case for the solar atmosphere, the frequency independent high-frequency asymptotic
gives the possibility to calculate in short wavelength approximation
the absorption coefficient
\be
A=\exp\left(-\int_0^{h_\mathrm{TR}}2k^{\prime\prime}(x)\md x \right)
\ee
and to evaluate the total heating power induced to the chromosphere 
by the high frequency part of the energy flux of the acoustic waves
\be
\int_0^{h_\mathrm{TR}}\mathcal{Q}_\zeta(x)\,\md x
=A \,q_\mathrm{ac}.
\ee
Taking in \Eqref{energy_per_hydrogen_1}
for singe ionization only, we obtain
\begin{align}
&
\overline{\varepsilon}
=\frac{T}{M^*}
\left(
\overline{\mathcal{N}}_\mathrm{tot}c_v
+\sum_{a}\overline{a}_a
\overline{\alpha}_{a}
\iota_{a}
\right),\\
&
\overline{w}
=\frac{T}{M^*}
\left[
\overline{\mathcal{N}}_\mathrm{tot}c_p
+\sum_{a}\overline{a}_a
\overline{\alpha}_{a}
\iota_{a}
\right],
\label{energy_per_mass_alpha}
\end{align}
where we repeat the definitions for the number or atoms 
and equilibrium electrons per hydrogen atom
\begin{align}
\overline{\mathcal{N}}_\mathrm{tot}=
\overline{\mathcal{N}}_a+\overline{\mathcal{N}}_e,
\quad \overline{\mathcal{N}}_a=\sum_a \overline{a}_a,\quad
\overline{\mathcal{N}}_e=\sum_a\overline{a}_a\overline{\alpha}_{a}.\nn
\end{align}

\subsection{General formula for the bulk viscosity expressed by 
complex polytropic index $\hat\gamma(\omega)$
and Drude approximation for $\hat\zeta(\omega)$}

\begin{figure}[ht]
\centering
\includegraphics[scale=0.5]{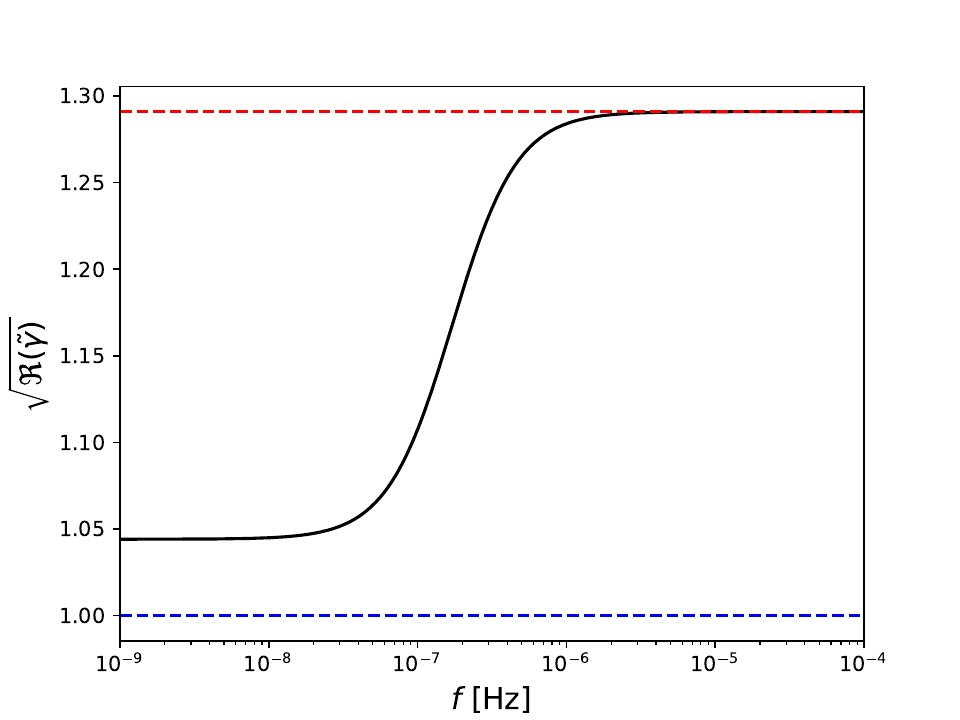}
\caption{Sound velocity ratio 
$c_\mathrm{s}/c_\mathrm{_N}=\sqrt{\gamma^\prime}$
as a function of frequency $f=\omega/2\pi$ according to the
calculations by the general formula \Eqref{complex_gamma_1}.
For high frequencies $\sqrt{\gamma^\prime(\infty)}=\sqrt{5/3}=\sqrt{\gamma_a}$ (upper horizontal dashed line)
this ratio reaches mono-atomic value. 
At low frequency $\sqrt{\gamma^\prime(0)}=\sqrt{\gamma(T,\rho)}\simeq1$,
the sound velocity is determined by the explicit thermodynamic expression
\Eqref{gamma_thermodynamic_1}.
The minimal value $\gamma_0\approx 1$ (the lower horizontal dashed line)
can be reached at $T\ll I_1$ for $\overline\alpha_1\simeq 1/2$.
}  
\label{fig:Re-gamma_1}
\end{figure}

\begin{figure}[ht]
\centering
\includegraphics[scale=0.5]{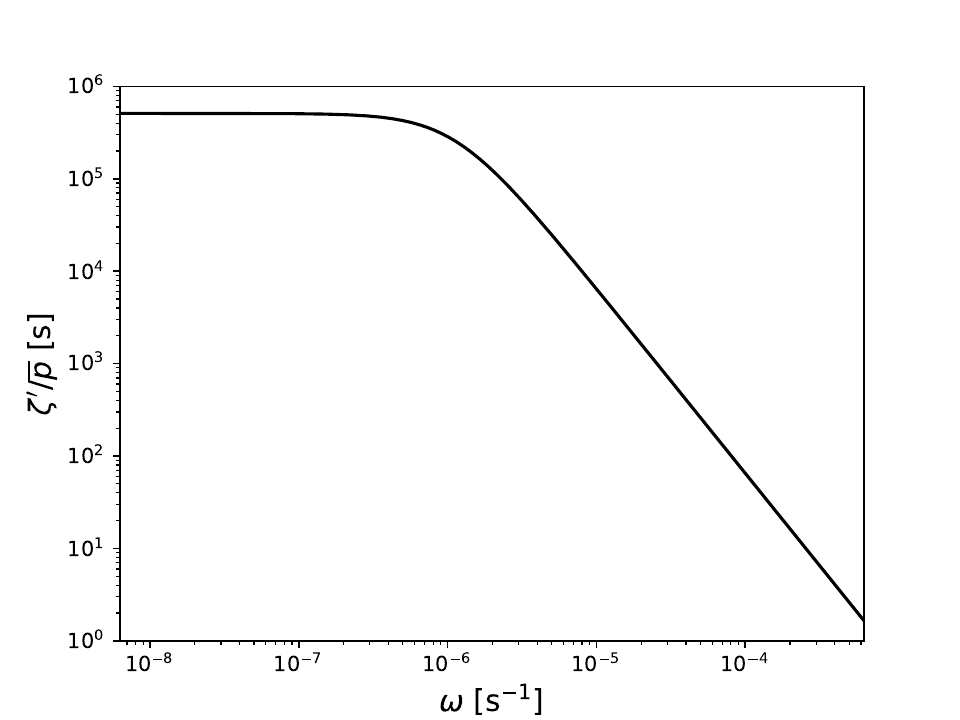}
\caption{Frequency dependence of the real (absorptive) part of the bulk viscosity
according to \Eqref{zeta_prime_1}.
On the ordinate is drawn $\zeta^\prime/\overline p$ with dimension of time, while
on the abscissa is the angular frequency $\omega$.
}  
\label{fig:zeta(omega)_1}
\end{figure}

Let us write again the formula for the wave vector with small
imaginary part created by bulk viscosity \Eqref{complex_k}, confer
Ref.~\cite[Eq.~(79.6) and Eq.~(81.9-10)]{LL6}
\begin{align}
\hat k(\omega) &= k^\prime+\im k^{\prime\prime}
=\left[1+\im\frac{\omega\,\zeta^\prime(\omega)}
{2\,\overline \rho \,c_\mathrm{s}^2(\omega)}\right]
\frac{\omega}{c_\mathrm{s}(\omega)}\\
&=\frac{\omega}{c_\mathrm{N}\sqrt{\hat\gamma(\omega)}}
\approx
\left[1-\im\frac{\gamma^{\prime\prime}}{2\gamma^\prime}\right]
\frac{\omega}{c_\mathrm{s}}, \label{k_prime_sec} \\
k^{\prime}(\omega)&=\Re(\hat k),\quad
k^{\prime\prime}(\omega)=\Im(\hat k),\quad
c_\mathrm{s}\equiv c_\mathrm{_N}\sqrt{\gamma^\prime}.
\end{align}
For $|\gamma^{\prime\prime}|\ll\gamma^{\prime}$, the comparison
gives the general formula for the bulk viscosity
expressed by the imaginary part of polytropic index
\be
\zeta^\prime(\omega)=-\frac{\gamma^{\prime\prime}(\omega)}{\omega}\,\overline p,
\qquad
\frac{\zeta^\prime(\omega)}{\overline\rho}
=-\frac{\gamma^{\prime\prime}(\omega)}{\omega}\,c_\mathrm{_N}^2,
\label{1_of_main_results}
\ee
which is one of the main results of the present study.
The sound velocity ratio $c_\mathrm{s}(\omega)/c_\mathrm{_N}=\sqrt{\gamma^\prime}$
as a function of the frequency $f=\omega/2\pi$ is depicted at \Fref{fig:Re-gamma_1}.
The ratio with dimension of time
\be
\frac{\zeta^\prime(\omega)}{\overline p}
=-\frac{\gamma^{\prime\prime}(\omega)}{\omega}
\label{zeta_prime_1}
\ee
is shown in \Fref{fig:zeta(omega)_1}.
The low frequency $\zeta_0$ limit has to be calculated 
for $\omega\tau\ll1$;
formally
\be
\zeta_0=-\overline p \lim_{\omega\rightarrow 0}
\frac{\gamma^{\prime\prime}(\omega)}{\omega}
=-\overline{p}\left.
\frac{\md\gamma^{\prime\prime}}{\md\omega}
\right|_{\omega=0}.
\label{zeta_0}
\ee
For very low frequencies 
$\gamma_0=\gamma(\omega=0)$
we have the explicit thermodynamical expression by the Jacobian 
\Eqref{gamma_thermodynamic_1}.
In such a way, the frequency dependent kinetic coefficient 
\be
\hat\zeta(\omega)=\im\,\frac{\hat\gamma(\omega)}{\omega}\,\overline{p}
\label{visc-cgama}
\ee
is obtained by the solution of the kinetic
equation applied to the pressure when we consider density fluctuations as a driving
perturbation.

The time-constant $\tau$ from the good working 
(see \Fref{fig:gama}) Mandelstam-Leontovich approximation \Eqref{M-L_1}
can be evaluated using \cite[Eq.~(81.7)]{LL6}
\be
\zeta_0=\tau\overline\rho(c_\infty^2-c_0^2),
\ee
which in our notations is
\be
\zeta_0=\tau \,\overline p\,(\gamma_\infty-\gamma_0).
\label{zeta_0}
\ee
At very small and very high frequencies, the imaginary part of the polytropic index is zero
\be
\gamma_0^{\prime\prime}=\gamma_\infty^{\prime\prime}=0,\qquad
\gamma_0^{\prime}=\gamma_0=\gamma(T,\rho),\quad 
\gamma_\infty^{\prime}=\gamma_a.
\ee
The comparison of this interpolation expression with the low-frequency result
gives the final result
for the Mandelstam-Leontovich time constant expressed by 
the complex frequency depended polytropic index
\begin{align}
&
\tau_0\equiv-\frac1{\gamma_\infty^\prime-\gamma_0^\prime}
\lim_{\omega\rightarrow 0}
\frac{\gamma^{\prime\prime}(\omega)}{\omega},
\label{tau_0}\\
&
\frac{\gamma_\infty^\prime-\gamma_0^\prime}{1/\tau_0}
=-\left.\frac{\md \gamma^{\prime\prime}(\omega)}{\md \omega}
\right|_{\omega=0},
\label{tau_determination_0}
\end{align}
where $\hat\gamma(\omega)=\gamma^\prime+\im\gamma^{\prime}$
is given by the sum \Eqref{complex_gamma_1} over small variations
of the ionization degrees $\hat\epsilon_a$ created by the 
density oscillations $\hat\epsilon_\rho\propto\e^{-\im\omega t}$.
The time constant $\tau_0 \approx 10.25$~days according to \Eqref{tau_0} which compared with the Mandelstam-Leontovich fitted $\tau_\mathrm{max} \approx 10.5$ days from \Eqref{Q_fit_1} and in \Fref{fig:recQ} have $\approx 2\%$ difference.
This excellent agreement is a consequence of the good working ML fit
plotted in \Fref{fig:gama}.


Using the so determined $\tau$, one can evaluate
\cite[Eq.~(81.9-10)]{LL6}
\begin{align}
&
k^{\prime\prime}(\omega\rightarrow 0) 
=\frac{(c_\infty^2-c_0^2)\tau\omega^2}{2c_0^3}
=\frac{\zeta_0}{2\rho c_0^3}\,\omega^2,
\label{k''low_1}\\
&
k^{\prime\prime}(\omega \rightarrow \infty)=\frac{c_\infty^2-c_0^2}{2\tau c_\infty^3}
=\frac{\zeta_0}
{2\tau^2\rho c_\infty^3}=\mathrm{const},
\label{k''_infty_1}
\end{align}
and also the Drude frequency dependence of the complex bulk viscosity
\cite[Eq.~(81.6)]{LL6}
\be
\hat{\zeta}(\omega)=\zeta^\prime+\im\zeta^{\prime\prime}
\approx\frac{\zeta_0}{1-\im\omega\tau},\quad
\zeta^\prime(\omega)\approx\frac{\zeta_0}{1+\omega^2\tau^2}.
\label{Drude_zeta_1}
\ee 

\begin{figure}[ht]
\centering
\includegraphics[scale=0.5]{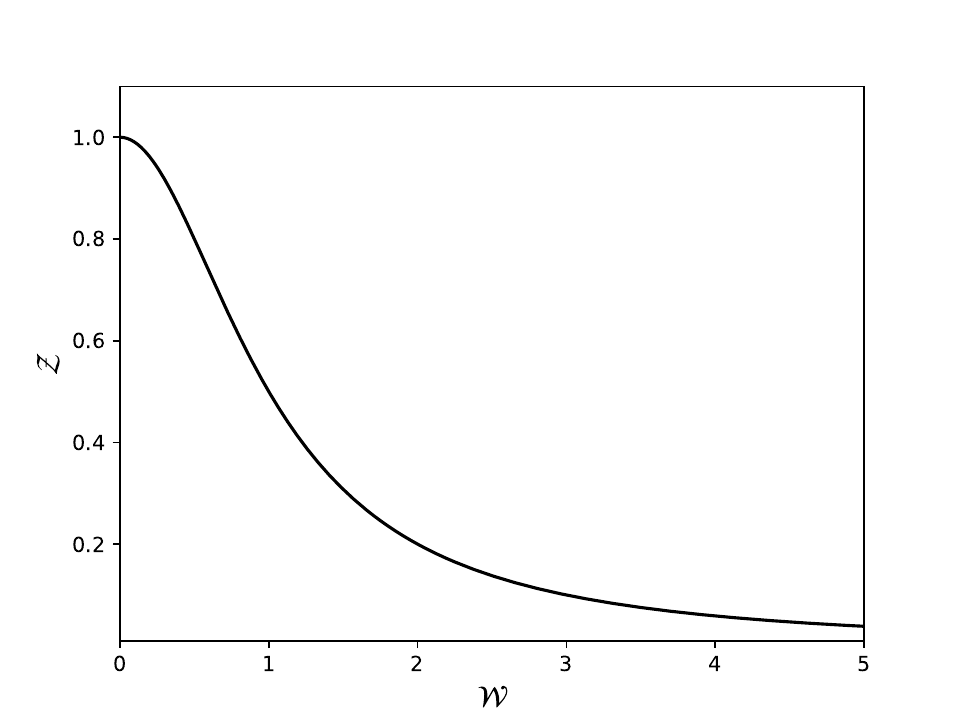}
\caption{
Frequency dependent dimensionless bulk viscosity 
$\mathcal{Z}=\zeta^\prime/\zeta_0$ as a
function of dimensionless frequency 
$\mathcal{W} \equiv \omega\tau$ 
according to \Eqref{Bash_Equation_1}.
The curve is undistinguished by its Drude fit.
}  
\label{fig:Drude_1}
\end{figure}

The frequency dependent bulk viscosity and its Mandelstam-Leontovich approximation
\be
\mathcal Z\equiv \zeta^\prime(\omega)/\zeta_0,\qquad \mathcal{W}=\omega\tau,
\qquad
1/\mathcal{Z}\approx 1+\mathcal{W}^2
\label{Drude_line}
\ee
obtained by solution of the kinetic equation 
\Eqref{Bash_Equation_1} are depicted in \Fref{fig:Drude_1}.
We consider that this plot will be useful for future experimental data processing.


Having reliable frequency dependence of the absorption part of the bulk viscosity 
\Eqref{Drude_zeta_1} with the routine calculation of $\hat\zeta(\omega)$
\Eqref{complex_gamma_1}, we have an alternative method
for determination of the ML relaxation time
\be
\tau_\infty\equiv\lim_{\omega\rightarrow\infty}
\left( \, \frac{1}{\omega}\sqrt{\frac{\zeta_0}{\zeta^\prime(\omega)}-1} \, \right)
\approx\sqrt{\frac{\zeta_0}{\omega^2\zeta^\prime(\omega)}}.
\label{tau_infty}
\ee
The numerical agreement of the so defined $\tau_\infty$
\Eqref{tau_infty}
with $\tau_0$ \Eqref{tau_0} and  
$\tau_\mathrm{max}$ from \Eqref{tau_max_1}
is the crucial test for the applicability of the ML approximation.
An agreement within several percent accuracy is a simple consequence
that at low temperature plasma, the number of a single type ions dominates.
In this case, imagine pure hydrogen, we have only one equation in
\Eqref{Bash_Equation_1} for the 
relative amplitude of oscillation hydrogen ionization degree $\hat\epsilon_1$.
Substitution in all general formulae reveals that for pure hydrogen,
plasma ML approximation is actually the exact solution.

\begin{figure}[ht]
\centering
\includegraphics[scale=0.5]{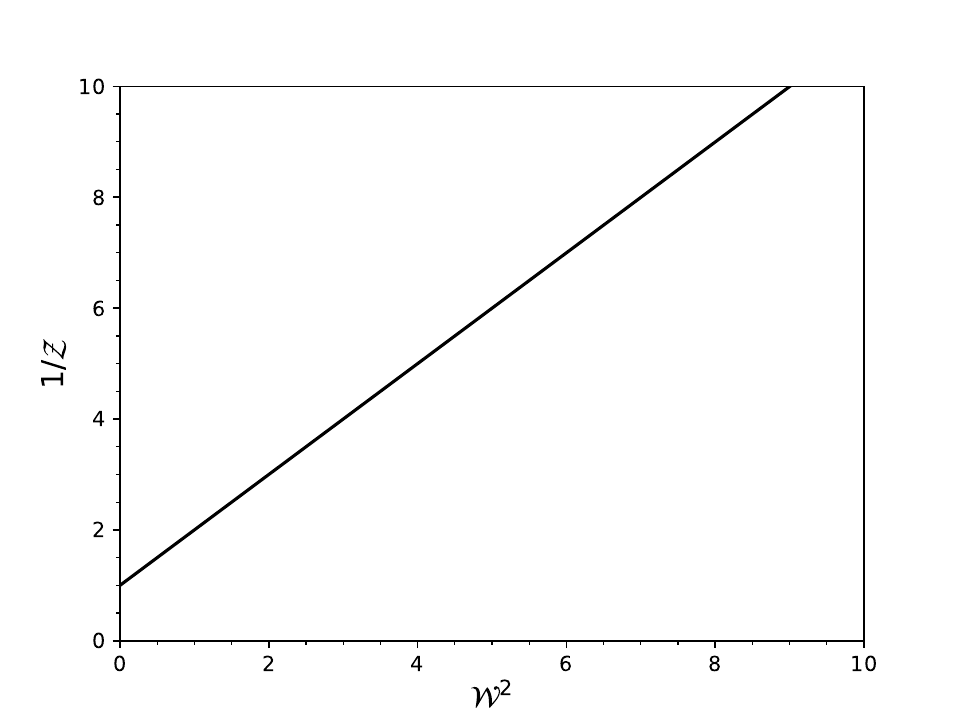}
\caption{The same results from the exact numerical solution of the
kinetic equation \Eqref{Bash_Equation_1}
represented in \Fref{fig:Drude_1} but in the plane
$1/\mathcal{Z}$ in the ordinate versus 
$\mathcal{W}^2$ in the abscissa.
For us, it was the most complicated method to draw with the computer pixel
accuracy the straight line $1/\mathcal{Z}=1+\mathcal{W}^2$ of the ML approximation.
The final result in the present study is that the ML approximation has an excellent
accuracy for cold plasmas where $T\ll I_a$.
}  
\label{fig:alphatau_1}
\end{figure}

That is why the so determined time-constant $\tau$ can be used 
even for the high frequency case $\omega\tau\gg1$ as it is 
represented in \Fref{fig:alphatau_1}.
This re-drawing in \Fref{fig:alphatau_1} of the Drude behavior as a straight line according to \Eqref{Drude_line} was done in order to observe any deviations of the calculated $\zeta(\omega)$ from its
ML approximation $\zeta^\mathrm{ML}(\omega)$.
We did not expect that this approximation will be so good.

\begin{figure}[ht]
\centering
\includegraphics[scale=0.5]{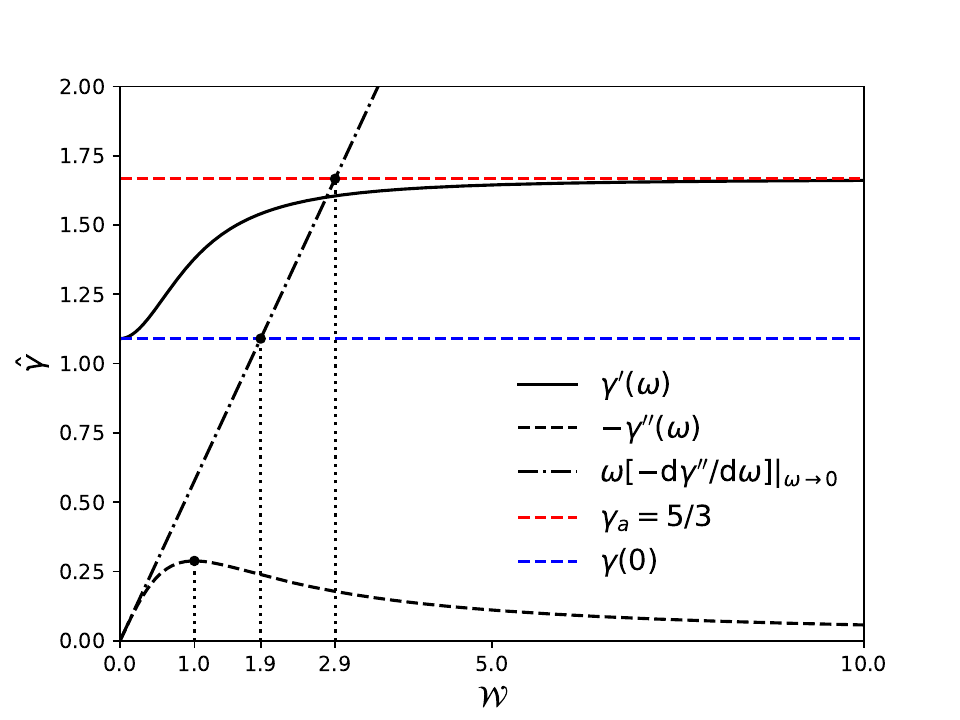}
\caption{Frequency dependence of the complex adiabatic index 
$\hat\gamma=\gamma^\prime+\im\gamma^{\prime\prime}$ 
as a function of the dimensionless frequency 
$\mathcal{W}=\omega\tau$. 
These functions together with the Argand plot
\Fref{fig:gama}
$\gamma^\prime$ versus $\gamma^{\prime\prime}$
gives the complete function $\hat\gamma(\omega\tau)$.
The real part $\gamma^\prime$ monotonously increases from
$\gamma_0$ to $\gamma_a=5/3$, while 
the absorptive part $\gamma^{\prime\prime}$ has a maximum at 
$\mathcal{W}_\mathrm{max}=\omega_\mathrm{max}\tau\approx 1$, 
confer \Fref{fig:recQ}: the accuracy of this approximation depends of the 
accuracy of ML fit according~\Fref{fig:gama}.
Its tangent at zero frequency 
$\omega [-\md \gamma^{\prime\prime}/\md\omega]|_{\omega \rightarrow 0}$
(the sloped dash-dotted line)
intercepts the horizontal lines $\gamma(0)$ and $\gamma_a$.
The difference of the abscissae of these intercepts is just $\Delta ( \omega\tau)=1.$
This is the tool to determine the time constant $\tau$ according to
\Eqref{tau_0} and \Eqref{tau_determination_0}.
}  
\label{fig:gammaW}
\end{figure}

The dimensionless complex function 
$\hat\gamma(\omega)=\gamma^\prime+\im\gamma^{\prime\prime}$
is the most important ingredient of our theoretical analysis of the bulk viscosity of the cold plasma.
In \Fref{fig:gammaW} its real and imaginary parts are depicted.
The real part $\gamma^\prime(\omega)$ monotonously increases from the thermodynamically calculated $\gamma_0$ to the monoatomic value $\gamma_a$.
The imaginary part $-\gamma^{\prime\prime}(\omega\tau)$ is negligible 
at very low and very high frequencies.
It has a maximum marked by a dot at the
frequency for which $\omega\tau=1$.
Dots at the $\gamma^\prime(\mathcal{W})$ dependence 
and the tangent at the $-\gamma^{\prime\prime}(\mathcal{W})$
illustrate the determination of the time-constant $\tau$ according to \Eqref{tau_determination_0}.

\begin{figure}[ht]
\centering
\includegraphics[scale=0.5]{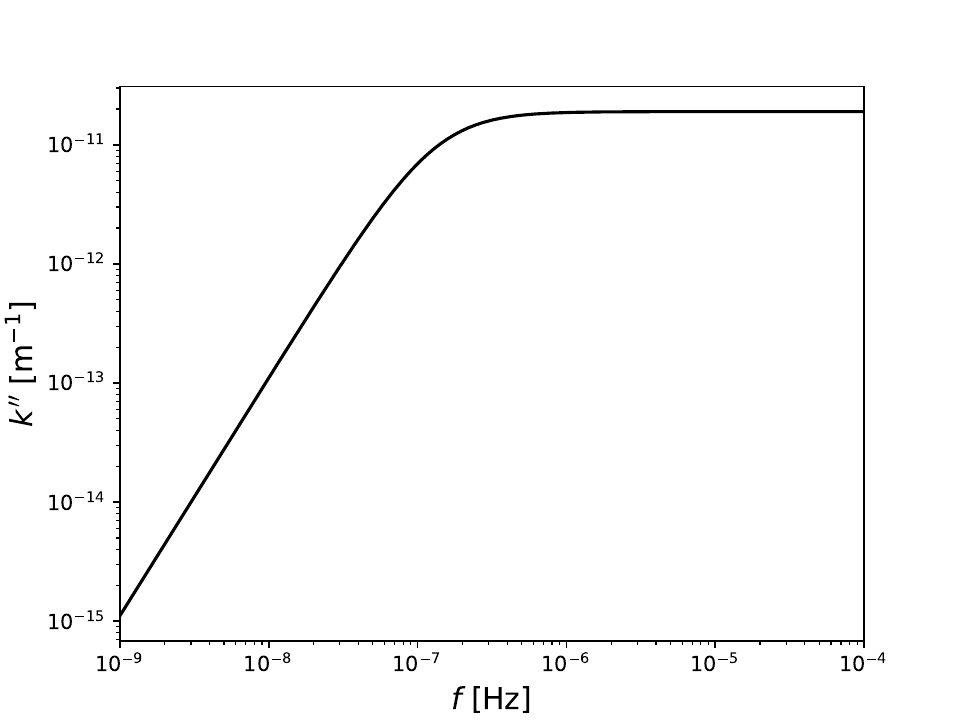}
\caption{Imaginary part of the wave-vector determining wave amplitude damping
$k^{\prime\prime}(f=\omega/2\pi)$
in a homogeneous fluid $\propto \e^{-k^{\prime\prime}x}$.
At low frequencies, the space damping rate $k^{\prime\prime}\propto\omega^2$
according to \Eqref{k''low_1}.
At high frequencies $\omega\tau\gg1$,  $k^{\prime\prime}$ reaches the high-frequency asymptotic $k_\infty^{\prime\prime}$ given by \Eqref{k''_infty_1}.
}  
\label{fig:k-sec}
\end{figure}

Perhaps in astrophysical problems, the bulk viscosity could be the most
important for calculation of acoustic heating by the space damping
rate $k^{\prime\prime}(f)$ as a function of the frequency.
This dependence is represented in logarithmic scale in \Fref{fig:k-sec}.
For extremely low frequencies we have power law dependence
$k^{\prime\prime}\propto f^2$ according to \Eqref{k''low_1}.
For higher frequencies, the damping rate is dispersion-less 
$k^{\prime\prime}\approx\mathrm{const}$
according to \Eqref{k''_infty_1}.
This is an important simplification:
in the frequency interval $\omega\in (1/\tau,\,\sqrt{\mathrm{P}_{\zeta/\eta}}/\tau )$
the spectral density of the heating power is proportional
to the spectral density of the acoustic waves.

Finally we have to compare the high frequency behavior of the bulk viscosity
$\zeta^\prime(\omega)\approx\zeta_0/(\omega\tau)^2$
with the shear one which has no frequency dependent dispersion.
They are equal at the critical frequency $f_c=\omega_c/2\pi$ for which
\begin{align}
\zeta^\prime(\omega_c)=\eta, \quad
\omega_c=\frac{1}{\tau}\sqrt{\mathrm{P}_{\zeta/\eta}}, \quad
\mathrm{P}_{\zeta/\eta}\equiv\frac{\zeta_0}{\eta}\gg 1.
\label{visc_Prandtl}
\end{align}
In order the short wavelength approximation to be applicable and the plasma to be considered as a homogeneous medium,
the wavelength must be smaller than the typical length $h$ of the system
\be
\lambda_c\equiv c_s/f_c\ll h.
\ee
For frequencies smaller than $f_c$,
the bulk viscosity dominates in the wave damping and acoustic heating.
In our illustrative example from the AL08 profile
\be
c_s \approx 9.24~\mathrm{km/s}, \quad
f_c \approx 35~\mathrm{mHz}, \quad
\lambda_c \approx 264~\mathrm{km}
\ee
at $h=1894$~km.
Our qualitative conclusion is that the bulk viscosity indispensably has to 
be included in the models for acoustic heating of the solar chromosphere~\cite{\AcousticHeating}
and models involving low temperature plasmas~\cite{Trelles:25}
when sound waves have significant spectral density for frequencies $\ll f_c$.
For the solar atmosphere such acoustic waves have already been found~\cite{Andretta:26}.

\section{Hydrogen-helium (alkali-noble) cocktail analytical solution for the thermodynamics and kinetics} \label{Sec_H-He}

In order to illustrate the work of the general scheme, here we analyze
the important special case of a cold cocktail of partially ionized hydrogen and neutral helium atoms for the solar chromosphere, i.e.
\begin{align}
&
\epsilon\equiv\epsilon_\mathrm{H}\equiv\epsilon_1,
\qquad
\epsilon_\mathrm{He}\equiv\epsilon_2=0,\\
&
\overline{a}_\mathrm{H}=1,
\qquad
\overline{a}_\mathrm{He}=0.1\,,\\
&
\rho=M^*n_\rho,\qquad M^*=M+\overline a_\mathrm{He}M_\mathrm{He},\\
&
1/\tau_\mathrm{_H}=1/\tau_1=\beta(T) n_e
=\beta n_\rho\overline{\alpha},\\
&
\overline{\mathcal{N}}_\mathrm{tot}=1+\overline{\alpha}
+\overline{a}_\mathrm{He},
\\
&
\overline{\alpha}(T)\equiv \overline{\alpha}_\mathrm{H}
=\dfrac{2}{1+\sqrt{4\dfrac{\overline{n}_\rho}{n_\mathrm{_S}}+1}},
\qquad
\overline{\alpha}_\mathrm{He}=0.
\end{align}
And for the pressure in the approximation of ideal gas we have
\be
p= n_\mathrm{tot}T= \mathcal{N}_\mathrm{tot}n_\rho T
=(1+\alpha+ \overline{a}_\mathrm{He})n_\rho T.
\label{pressure_init}
\ee
The Saha equation can be written as
\begin{align}
& \alpha(\rho,T) = f(\nu) \equiv \alpha_\mathrm{H} \equiv\dfrac{2}{1+\sqrt{1+4\nu}},
 \label{alpha_explicit} \\
& \nu\equiv\dfrac{n_\rho}{n_\mathrm{_S}} = \frac{1-\alpha}{\alpha^2}, 
\qquad
\dfrac1{\sqrt{1+4\nu}}=\dfrac{\alpha}{2-\alpha}
\label{nu_equiv}
\\
& n_\mathrm{_S}(T) \equiv n_q \e^{-\iota},
\quad
n_q(T) = \left(\dfrac{mT}{2\pi\hbar^2}\right)^{\!\!3/2},
\quad
\iota\equiv\dfrac{I}{T}, \nn
\end{align}
where $m$ is the electron mass and $n_\rho=\rho/M^*$,
Its solution for the hydrogen degree of ionization 
$\overline\alpha(T,\overline{n}_\rho)$ 
is given in Ref.~\cite{PhysA}.
Actually, this is an applicable approximation for the solar chromosphere but the
derived formulas are applicable for any alkali-noble cocktail, Na-Ne plasma or arbitrary cold pure element plasma.\\

\subsection{Thermodynamics of the cold binary plasma}

For the enthalpy $w$ 
and internal energy $\varepsilon$
per unit mass we have \cite[Eqs.~(11-12), Eq.~(6)]{ApJ:21,ApJ:24}
\begin{align}
w(\rho,T)&=\dfrac{1}{M^*}\left(c_p\mathcal{N}_\mathrm{tot}T+I\alpha\right),
\label{enthalpy_per_unit_mass}
\qquad
c_p=\dfrac52,\\
\varepsilon(\rho,T)&=\dfrac{1}{M^*}\left(c_v\mathcal{N}_\mathrm{tot}T+I\alpha\right),
\qquad
c_v=\dfrac32.
\end{align}
Here we wish to emphasize that  magnetic field has zero influence 
on the thermodynamics of the solar atmosphere which is a classical gas.
This statement is known as Bohr--van~Leeuwen theorem~\cite{VanLeeuwen}.
For the derivatives we have
\begin{align}
f^\prime(\nu) & =  \md_\nu f(\nu)
=-\dfrac4{\left(1+\sqrt{1+4\nu}\right)^{\!2}\sqrt{1+4\nu}}
\nn \\
& = \dfrac{-f^2(\nu)}{\sqrt{1+4\nu}}
=\dfrac{-f^3}{2-f}=-\dfrac{\alpha^3}{2-\alpha}=\dfrac{\md \alpha}{\md \nu},
\\
\nu f^\prime(\nu) & = -\dfrac{(1-\alpha)\alpha}{2-\alpha}=-\mathcal{D},
\quad \mathcal{D}\equiv \dfrac{(1-\alpha)\alpha}{2-\alpha},
\label{cal_D}
\\
\left(\dfrac{\partial \alpha}{\partial T}\right)_{\!\!\rho} & = -\dfrac{(c_v+\iota)}{T}\nu f^\prime(\nu)
=\dfrac{(c_v+\iota)}{T}\dfrac{(1-\alpha)\alpha}{2-\alpha}\nn\\
&=\dfrac{(c_v+\iota)}{T}\mathcal{D},
\\
\left(\dfrac{\partial \alpha}{\partial \rho}\right)_{\!T} & = \nu f^\prime(\nu)/(M^*n_{\rho})
=-\dfrac{1}{M^*n_{\rho}} \dfrac{(1-\alpha)\alpha}{2-\alpha}\nn\\
& =-\dfrac{1}{M^*n_{\rho}} \mathcal{D} 
 = -\dfrac{\mathcal{D}}{\rho},\\
T\left(\dfrac{\partial \alpha}{\partial T}\right)_{\!\!\rho} & =
(c_v+\iota)\mathcal{D}, \qquad 
\rho\left(\dfrac{\partial \alpha}{\partial \rho}\right)_{\!\!T}  = -\mathcal{D}.
\end{align}
In such a way we obtain for the heat capacity per constant volume the explicit expression
\begin{align}
\mathcal{C}_v & \equiv \left(\dfrac{\partial \varepsilon}{\partial T}\right)_{\!\!\rho}
=\dfrac{1}{M^*}\left[c_v\mathcal{N}_\mathrm{tot}+ (c_v +\iota)^2  \mathcal{D}\right]
\\
& = \dfrac{1}{M^*}
\left[c_v(1+\alpha+\overline{a}_\mathrm{He})
+ (c_v +\iota)^2\dfrac{(1-\alpha)\alpha}{2-\alpha}\right]\\
& = T\left(\dfrac{\partial s}{\partial T}\right)_{\!\!\!\rho},
\label{C_v_general}
\end{align}
where $s$ is the entropy per unit mass.

For derivatives of the pressure \Eqref{pressure_init} we have
\begin{align}
\left(\dfrac{\partial p}{\partial T}\right)_{\!\!\rho}
&=\left[\mathcal{N}_\mathrm{tot}+(c_v+\iota)\mathcal{D}\right]n_\rho 
\label{dp/dT}
\\
&=\left[ (1+\alpha+\overline{a}_\mathrm{He})+(c_v+\iota)
\dfrac{(1-\alpha)\alpha}{2-\alpha}\right]n_\rho,\nn \\
\left(\dfrac{\partial p}{\partial \rho}\right)_{\!\!T}
&=\dfrac{T}{M^*}\left[\mathcal{N}_\mathrm{tot}-\mathcal{D}\right] \label{dp/d_rho}\\
&=\dfrac{T}{M^*}
\left[ (1+\alpha+\overline{a}_\mathrm{He})-
\dfrac{(1-\alpha)\alpha}{2-\alpha}\right]. \nn
\end{align}
The difference between the heat capacities 
can be obtained by the well-known formula~\cite[Eq.~(16.10)]{LL5} in which $\mathcal{V}\equiv1/\rho$
\begin{align}
\Delta\mathcal{C} & \equiv
\mathcal{C}_p-\mathcal{C}_v
=-T\dfrac{(\partial p/\partial T)_\mathcal{V}^2}
{(\partial p/\partial \mathcal{V})_T}
=\dfrac{T}{\rho^2}
\dfrac{(\partial p/\partial T)_\rho^2}
{(\partial p/\partial \rho)_T}\nn\\
&=\dfrac{\left[\mathcal{N}_\mathrm{tot}+(c_v+\iota)\mathcal{D}\right]^2}
{M^*\left[\mathcal{N}_\mathrm{tot}-\mathcal{D}\right]}
\label{Delta_C}\\
&=\dfrac{\left[ (1+\alpha+\overline{a}_\mathrm{He})+(c_v+\iota)
\dfrac{(1-\alpha)\alpha}{2-\alpha}\right]^2}
{M^*\left[ (1+\alpha+\overline{a}_\mathrm{He})-
\dfrac{(1-\alpha)\alpha}{2-\alpha}\right]}. \nn
\end{align}
Then for the heat capacity per unit mass at constant pressure we obtain
\begin{align}
\mathcal{C}_p=&
\dfrac{\left[c_v\mathcal{N}_\mathrm{tot}+ (c_v +\iota)^2  \mathcal{D}\right]}{M^*}
+\dfrac{\left[\mathcal{N}_\mathrm{tot}+(c_v+\iota)\mathcal{D}\right]^2}
{M^*\left[\mathcal{N}_\mathrm{tot}-\mathcal{D}\right]} \nn \\
=&\dfrac{\left[c_p\mathcal{N}_\mathrm{tot}+(c_vc_p+2c_p\iota+\iota^2)\mathcal{D}\right]\mathcal{N}_\mathrm{tot}}{(\mathcal{N}_\mathrm{tot}-\mathcal{D})M^*}\label{C_p}
\\
= &\dfrac{1}{M^*}
\left[c_v(1+\alpha+\overline{a}_\mathrm{He})
+ (c_v +\iota)^2\frac{(1-\alpha)\alpha}{2-\alpha}\right]\nn\\
&+
\dfrac{\left[ (1+\alpha+\overline{a}_\mathrm{He})+(c_v+\iota)
\dfrac{(1-\alpha)\alpha}{2-\alpha}\right]^2}
{M^*\left[ (1+\alpha+\overline{a}_\mathrm{He})-
\dfrac{(1-\alpha)\alpha}{2-\alpha}\right]}\\
& = T\left(\dfrac{\partial s}{\partial T}\right)_{\!\!p}
=\mathcal{C}_v+\Delta\mathcal{C}.
\end{align}
These results generalize the previously obtained heat capacities
for pure hydrogen~\citep{ApJ:21}.  

Analogously for the derivatives of enthalpy we have
\begin{align}
\left(\dfrac{\partial w}{\partial T}\right)_{\!\!\rho}
& =\dfrac1{M^*}
\left[c_p\mathcal{N}_\mathrm{tot}+(c_v+\iota)(c_p+\iota)\mathcal{D}\right]\\
&=\left[ (1+\alpha+\overline{a}_\mathrm{He})c_p+(c_v+\iota)(c_p+\iota)
\dfrac{(1-\alpha)\alpha}{2-\alpha}\right], \nn \\
\left(\dfrac{\partial w}{\partial \rho}\right)_{\!\!T}
&=-\dfrac{T}{M^*}(c_p+\iota)\dfrac{\mathcal{D}}{\rho} \\
&=-\dfrac{T}{M^*}\dfrac{1}{M^*n_\rho}
\left[ (c_p+\iota)
\dfrac{(1-\alpha)\alpha}{2-\alpha}\right].\nn
\end{align}

Then the Jacobian~\cite[Eq.~(9)]{ApJ:24} with dimension of reciprocal mass
\be
 \mathcal{J} \equiv 
\frac{\partial (w,p)}{\partial (T,\rho)} 
= 
\left(\frac{\partial w}{\partial T}\right)_{\!\! \rho}
\left(\frac{\partial p}{\partial \rho}\right)_{\!\! T}
-\left(\frac{\partial w}{\partial \rho}\right)_{\!\! T}
\left(\frac{\partial p}{\partial T}\right)_{\!\! \rho}
\label{Jacobian_up} 
\ee
after some algebra reads
\begin{widetext}
\be
\mathcal{J} =\frac{T\mathcal{N}_\mathrm{tot}}{(M^*)^2}
\left\{c_p\mathcal{N}_\mathrm{tot}+
\left[(c_v+\iota)(c_p+\iota)+\iota\right]
\mathcal{D}
\right\}
=\frac{(1 \! + \! \alpha \! + \! \overline{a}_\mathrm{He})T}{(M^*)^2}
\left[ c_p(1 \! +\! \alpha \! + \! \overline{a}_\mathrm{He})
+
(c_vc_p+2c_p \iota+\iota^2)
\frac{(1-\alpha)\alpha}{2-\alpha}
\right].
\label{Jacobian_analytical}
\ee
\end{widetext}
In such a way we have obtained that for the 2-component cocktail
\be
\mathcal{J}=\frac{T}{M^*}
(\mathcal{N}_\mathrm{tot}-\mathcal{D})\mathcal{C}_p.
\label{simple_J}
\ee
Representing the sound velocity at evanescent frequency when the
ionization degree follows its equilibrium values determined by the Saha equation
we have
\begin{eqnarray}
c_0^2\equiv\left(\frac{\partial p}{\partial \rho}\right)_{\!\!s}=\gamma_0\frac{p}{\rho}.
\end{eqnarray}
Using the general result~\cite[Eq.~(9)]{ApJ:24}
\be
\gamma_0=\frac{\rho}{p}\cdot\frac{\mathcal{J}}{\mathcal{C}_p}
\ee
after substitution $\mathcal{J}$ from \Eqref{simple_J} we arrive at
\begin{eqnarray}
\gamma_0&=&\left(1-\dfrac{\mathcal{D}}{\mathcal{N}_\mathrm{tot}}\right)
\dfrac{\mathcal{C}_p}{\mathcal{C}_v}
\label{gama_simple}
\\
&=&\left[1-\dfrac{(1-\alpha)\alpha}{(2-\alpha)(1+\alpha+\overline{a}_\mathrm{He})}
\right]
\dfrac{\mathcal{C}_p}{\mathcal{C}_v}. \nn
\end{eqnarray}
This illustrates how the ionization recombination processes break
the general relations derived for constant chemical compounds.
For the solar plasma it is not so strongly expressed but for the interstellar plasma where $\iota=I/T\gg1$ it is possible $\iota^2\mathcal{D}$ to dominate, hence
\begin{eqnarray}
M^*\mathcal{C}_v&\approx&\iota^2\mathcal{D}\gg1,\\
M^*\mathcal{C}_p&\approx&
\dfrac{\iota^2 \mathcal{D}}{1-\mathcal{D}/\mathcal{N}_\mathrm{tot}}\gg1,
\end{eqnarray}
confer Ref~\cite[Fig.~5]{PhysA}.
In the opposite case of negligible influence of the ionization processes
$\iota\ll1$, we have the test for the programming of the heat capacities
\be
M^*\mathcal{C}_v\approx c_v\mathcal{N}_\mathrm{tot},\;
M^*\mathcal{C}_p\approx c_p\mathcal{N}_\mathrm{tot},\;
M^*(\mathcal{C}_p-\mathcal{C}_v)=\mathcal{N}_\mathrm{tot}.
\ee

The general formula for pure
hydrogen plasma was solved in the early stages of the present study~\cite[Eq.~(96)]{PhysA} and \cite[Eq.~(17)]{ApJ:21}
\be
c_0^2=\frac{2c_p+(1-\alpha)\alpha(c_p+\iota)^2}{2c_v+(1-\alpha)\alpha\left[(c_v+\iota)^2+c_v\right]}\frac{\bar{p}}{\bar{\rho}},
\label{c0}
\ee
where $\iota \equiv I/T$, $I=13.6$~eV is the hydrogen ionization potential,
$\alpha\equiv \overline\alpha$ is the equilibrium ionization degree
\begin{align}&
\alpha\equiv\frac{n_e}{n_p+n_0}=\frac{n_e}{n_\rho} \in(0,1)\,,\\&
\label{nrho}
n_e=n_p=\alpha n_\rho,\quad n_0=(1-\alpha)\,n_\rho, \quad \\&
n_\mathrm{tot}=n_e+n_p+n_0=(1+\alpha)\,n_\rho,
\end{align}
where $n_e$, $n_p$ and $n_0$ are the concentrations of electrons, protons and neutral atoms respectively, and we denote the sum of the concentrations of the protons and the neutrals by $n_\rho=n_0+n_p$.
This formula can be generalized for arbitrary
low temperature cocktail of alkaline metal and noble gas~\cite[Eq.~(B1)]{ApJ:24},
(for example Na-Ne plasma)
\be
\gamma_0=\frac{c_p
[2+(2-\alpha)\overline{a}_\mathrm{nbl}]
+(1-\alpha)\alpha(c_p+\iota)^2}
{c_v
[2+(2 \! - \! \alpha)\overline{a}_\mathrm{nbl}]
+(1 \! -\! \alpha)\alpha\left[(c_v+\iota)^2+c_v\right]},
\label{c0_cocktail_1}
\ee
where
\be
\overline{a}_\mathrm{nbl}
=\frac{n_\mathrm{nbl}}{n_\mathrm{alk}},
\quad \alpha \equiv \overline{\alpha}=\overline r_{1,\,\mathrm{alk}},
\qquad
I=I_{1,\,\mathrm{alk}}.\nn
\ee

\subsection{Kinetics of the cold alkali-noble plasma}

In this special case the general \Eqref{Bash_Equation_1} is reduced to 
a simple linear equation for $\epsilon$ having the solution
\be
\epsilon=\dfrac{\dfrac{\iota}{c_v} \epsilon_\rho}{
\dfrac{1-\im\omega\tau_{1}\overline{\alpha}}{1-\overline{\alpha}}
+\dfrac{\overline{\alpha}}{\overline{\mathcal{N}}_\mathrm{tot}}
\left[\left(\dfrac{\iota}{c_v}+2\right)\!\iota+c_v+\dfrac{1}{\overline{c}_e}
\right]}.
\ee
The substitution of this solution in the final expression for the generalized
polytropic index \Eqref{general_gamma} gives
\begin{widetext}
\begin{align}
\label{gamma_alkaline_noble}
\hat\gamma(\omega) & = \frac{c_p}{c_v}
-\dfrac{\left(\dfrac{\iota}{c_v}\right)^{\!2}(1-\overline{\alpha})\,\overline{\alpha}}
{ (1-\im\omega\tau_\mathrm{_H} \overline{\alpha})(1+\overline{\alpha}
+\overline{a}_\mathrm{He})
+\left[\left(\dfrac{\iota}{c_v}+2\right)\!\iota+c_v+
\dfrac{1+\overline{\alpha}
+\overline{a}_\mathrm{He}}{\overline{\alpha}}
\right](1-\overline{\alpha})\,\overline{\alpha}
}\\
& = \gamma_a
-\dfrac{(1-\overline{\alpha})\,\overline{\alpha} \, \iota^2/c_v}
{ (1-\im\omega\tau_\mathrm{_H} \overline{\alpha}) c_v \overline{\mathcal{N}}_\mathrm{tot}
+\left[(c_v+\iota)^2\overline{\alpha} + 
c_v \overline{\mathcal{N}}_\mathrm{tot}
\right](1-\overline{\alpha})
} 
\label{gamma_not}
\\
& = \frac{c_p}{c_v}
-\dfrac{\left(\dfrac{\iota}{c_v}\right)^{\!2}(1-\overline{\alpha})\,\overline{\alpha}}
{ (1-\im\omega\tau_\mathrm{_H} \overline{\alpha})(1+\overline{\alpha}
+\overline{a}_\mathrm{He})
+\left[
\dfrac{(c_v+\iota)^2}{c_v}+
\dfrac{1+\overline{\alpha}+\overline{a}_\mathrm{He}}{\overline{\alpha}}
\right](1-\overline{\alpha})\,\overline{\alpha}
}.
\end{align}
\end{widetext}
This solution \Eqref{gamma_not} can be rewritten as
\begin{align}&
\hat\gamma(\omega)=\gamma_\infty-\frac{\mathcal{B}}
{\mathcal{A}-\im\omega\mathcal{T}},
\label{notions_A_B}\\
&
\mathcal{A}\equiv
(2-\overline{\alpha})c_v \overline{\mathcal{N}}_\mathrm{tot}
+(c_v+\iota)^2(1-\overline\alpha)\overline\alpha, \\
&
\mathcal{B}=(1-\overline{\alpha})\,\overline{\alpha} \, \iota^2/c_v, \\
&
\mathcal{T}
\equiv\tau_\mathrm{_H} c_v \overline\alpha \overline{\mathcal{N}}_\mathrm{tot},
\end{align}
and in these notations
\be
\gamma^{\prime\prime}
=-\frac{\mathcal{B}\,\omega\mathcal{T}}
{\mathcal{A}^2
+(\omega\mathcal{T})^2}, \qquad
-\left.\frac{\gamma^{\prime\prime}(\omega)}{\omega}\right|_{\omega\rightarrow 0}
=\frac{\mathcal{B}\mathcal{T}}
{\mathcal{A}^2}.
\ee
It is instructive to note that 
\begin{align}
\mathcal{\overline{N}}_\mathrm{tot} & =
1+\overline{a}_\mathrm{He}+\overline\alpha \equiv 
\overline{a}_\mathrm{H} +\overline{a}_\mathrm{He} + \overline{a}_e \nn\\
&
= \sum_{a}^{a_\mathrm{max}} \overline{a}_a + \overline{a}_e,
\quad \overline{a}_\mathrm{H}=1,\quad
\overline{a}_e=\overline\alpha,
\end{align}
clearly giving the recipe to include more noble gases in the cocktail.
Additionally, from \Eqref{notions_A_B} we have
\be
\gamma_\infty^\prime-\gamma_0^\prime
=\gamma_a-\gamma(\omega=0)=\mathcal{B}/\mathcal{A}
\ee
and the substitution in \Eqref{tau_0} gives for the Mandelstam-Leontovich
time-constant
\begin{align}
\tau&=
\frac{\mathcal{T}}{\mathcal{A}}
=\frac{\overline{\alpha} c_v\overline{\mathcal{N}}_\mathrm{tot}}
{\mathcal{A} } \tau_\mathrm{_H}\\
&=\frac{c_v \overline{\alpha}\overline{\mathcal{N}}_\mathrm{tot}}
{(2-\overline{\alpha})c_v \mathcal{N}_\mathrm{tot}
+(c_v+\iota)^2(1-\overline\alpha)\overline\alpha }\tau_\mathrm{_H}\\
&
=\frac{(1\!+\!\overline{a}_\mathrm{He}\!+\!\overline\alpha)
\overline{\alpha} c_v }
{(2\!-\!\overline{\alpha})(1\!+\!\overline{a}_\mathrm{He}\!+\!\overline\alpha)c_v 
+(c_v\!+\!\iota)^2(1\!-\!\overline\alpha)\overline\alpha}\tau_\mathrm{_H}.
\label{ML_time}
\end{align}
This is in agreement with the result for pure hydrogen plasma ~\cite[Eq.~(113)]{PhysA}
derived in a different way.
The time constant substituted in \Eqref{zeta_0} gives, cf.~\cite[Eq.~{108}]{PhysA}
\begin{align}
\zeta_0&
=\frac{\overline p\mathcal{B}\mathcal{T}}{\mathcal{A}^2}
=\frac{\overline{p}\mathcal{B}\overline{\alpha} 
c_v\overline{\mathcal{N}}_\mathrm{tot}\tau_\mathrm{_H}}
{\mathcal{A}^2}
\label{zeta_zero}\\
&=\frac{(1-\overline{\alpha})\,\overline{\alpha} \, \iota^2
(1+\overline{a}_\mathrm{He}+\overline\alpha) \overline{p}
\tau_\mathrm{_H}}
{[(2-\overline{\alpha})(1+\overline{a}_\mathrm{He}+\overline\alpha)c_v 
+(c_v+\iota)^2(1-\overline\alpha)\overline\alpha]^2}.\nn
\end{align}
For the frequency dependent sound velocity $c_\omega$ 
the ML theory gives the real part of 
of the complex phase velocity~\cite[Eq.~(81.8)]{LL6}
\begin{align}
&c_\omega\approx\Re(\hat c_\mathrm{phase})=\frac{\omega}{\Re( \hat k)}=\Re \left(
\sqrt{\frac{c_0^2-\im\omega c_\infty^2}{1-\im\omega\tau}}
\right)
,\\
&\hat k=\omega
\sqrt{\frac{1-\im\omega\tau}{c_0^2-\im\omega c_\infty^2} }.
\label{c_omega}
\end{align}

Finally the important for us high-frequency dispersion-less damping
according to \Eqref{k''_infty_1}
\begin{align}
2k_\infty^{\prime\prime}&\equiv
2k^{\prime\prime}(\omega \rightarrow \infty)
=\frac{1}{\tau c_\infty}\frac{\gamma_a-\gamma_0}{\gamma_a}
=\frac{1}{\tau c_\infty}\frac{\mathcal{B}}{\gamma_a\mathcal{A}}\\
& =\frac{\beta(T)n_\rho\iota^2(1-\overline\alpha)\overline\alpha}
{ (1+\overline\alpha+\overline a_\mathrm{He}) c_\infty c_p c_v}\\
&
= \frac{(\gamma_a-1)^2 (1-\overline\alpha)\overline\alpha \iota^2}{\gamma_a \overline{ \mathcal{N}}_\mathrm{tot}} 
\frac{1}{\tau_\alpha c_\infty},  \\
&
\rightarrow \sum_a \frac{I_a^2}{c_p c_v T^2}
\frac{\langle \sigma_a v \rangle }{c_\infty}
\frac{\overline n_e \overline n_{0,a}}{\overline n_\mathrm{tot}}\equiv \frac{1}{l}, \quad 
\tau_\alpha = \frac{\beta}{n_\rho},
\label{damping_inf}\\
c_\infty&=\sqrt{\frac{\gamma_a \overline p}{\overline \rho}},
\quad
\gamma_a=\gamma_\infty=\frac53,
\quad 
c_p=\gamma_a c_v,
\quad
c_v=\frac32.\nn
\end{align}
See also the result for pure hydrogen plasma~\cite[Eq.~(112)]{PhysA}
The last expression akin of the Clausius concept for the mean free path $l \sigma n \sim 1$.
The reciprocal mean free path $1/l$ is additive 
and sum of all kind scattering centers, in our case of all kind neutral atoms.

In the general case we have $\hat k(\omega)$ expressed by 
\Eqref{hat_k} with $\hat\gamma(\omega)$ from 
\Eqref{notions_A_B}.
In such a way, our general conclusion is that in
alkaline-noble gas approximation we have analytical expression
for $\zeta_0$ and $\tau$, and the Mandelstam-Leontovich
formula for $\zeta(\omega)$ gives the exact frequency 
dependence.

\subsection{Special cases analysis}

\begin{itemize}
\item Completely ionized hydrogen plasma $\overline{\alpha}=1$: 
$\hat\gamma(\omega)=\gamma_a$.
\item  Atomic hydrogen only $\overline{\alpha}=0$: 
$\hat\gamma(\omega)=\gamma_a$, i.e. for monoatomic gases the bulk viscosity is zero~\cite[Eq.~(8.16)]{LL10}.
\item Very high frequencies $\omega\rightarrow\infty$:
$\hat\gamma(\omega)=\gamma_a$ and $\zeta^{\prime}=0$ again because
the influence of the ionization-recombination processes in this 
$\omega\tau\gg1$ case is negligible.
\item In the opposite case of evanescent frequencies at formally $\omega=0$,
we reproduce after some algebra $\hat{\gamma}(\omega = 0)=\gamma_0$
the thermodynamic result \Eqref{c0_cocktail_1}~\cite[Eq.~(B1)]{ApJ:24},
with $\overline{a}_\mathrm{nbl} \equiv \overline{a}_\mathrm{He}$ and 
$\alpha \equiv \overline{\alpha}$.
\item Interstellar plasma for which $T\ll I$ at significant 
$(1-\overline{\alpha})\overline{\alpha}$, formally at 
$(1-\overline{\alpha})\overline{\alpha}\iota^2\gg1$:
we obtain
$\gamma_0\approx \gamma_a-1/c_v=1$, $c_0=c_\mathrm{N}$ and $c_0^2=\overline{p}/\overline{\rho}$.
This formula suggested initially by Newton is an important test for the
general program for numerical calculation of bulk viscosity of cold plasma.
\item As a function of the frequency $\omega$ 
\Eqref{gamma_alkaline_noble} gives a simple algebraic expression,
linear function divided by another linear function and taking into account the 
limit cases at low and high frequencies, we conclude that
in this singly ionized atomic approximation we arrive at the
Mandelstam-Leontovich
formula \Eqref{M-L_1}, which we have derived in the present study by the microscopic kinetic approach.
\item
Finally, knowing the general relation between the bulk viscosity and the generalized polytropic index \Eqref{1_of_main_results}, we can determine the 
relaxation time $\tau$ of the ML theory.
For example, comparing the high frequency behavior \Eqref{tau_infty} 
with the low frequency approximation \Eqref{tau_0}, we again arrive at
\Eqref{ML_time}.
\item Having a reliable approximation for low temperature hydrogen-helium
cocktail, we can address some practical for plasma physics problems as
the long standing problem of heating of the solar chromosphere, for example.
\end{itemize} 

\subsection{Solar chromospheric profiles of the main notions and damping of magneto-hydrodynamic waves}

Using the semi-empirical height dependent profiles of the temperature $T^\prime$ and mass density $\rho$~\cite[Model C7]{Avrett:08}~(AL08), the main introduced notions are calculated and analyzed in this subsection.

\begin{figure}[ht]
\centering
\includegraphics[scale=0.5]{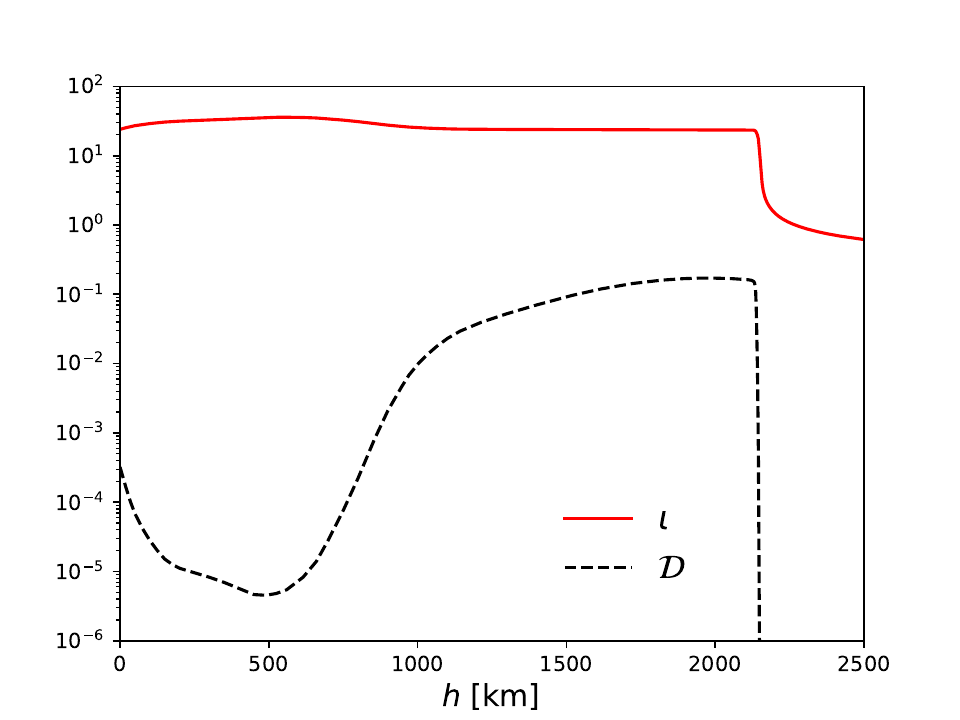}
\caption{The height dependence of dimensionless functions $\lg{[\iota(h) \equiv I/T(h)]}$ (solid line) and $\lg\mathcal{D}(h)$ (dashed line)
\Eqref{cal_D} which is via Ref.~\cite[Model C7]{Avrett:08}
in the ordinate for heights $h \equiv x <2.5$~Mm in the abscissa.
The function $\mathcal{D}$ is the main ingredient of all analytical results in which
the ionization-recombination processes are relevant.
For comparison, in the same logarithmic ordinate
the profile of the dimensionless variable $\iota(h)\gg1$ is given.
The abrupt change of both variables is physically in the solar transition region and it is beyond the scope of the present study.
}  
\label{fig:D}
\end{figure}

First, the height profile of the function $\mathcal{D}(h)$
is depicted in \Fref{fig:D} below the profile of $\iota(h)$. 
Both these notions are central ingredients in our analytical results.
As $\mathcal{D} \rightarrow 0$ in the solar transition region, $\zeta \rightarrow 0$ and $\iota < 1$ in the coronal conditions beyond the current study.

\begin{figure}[ht]
\centering
\includegraphics[scale=0.5]{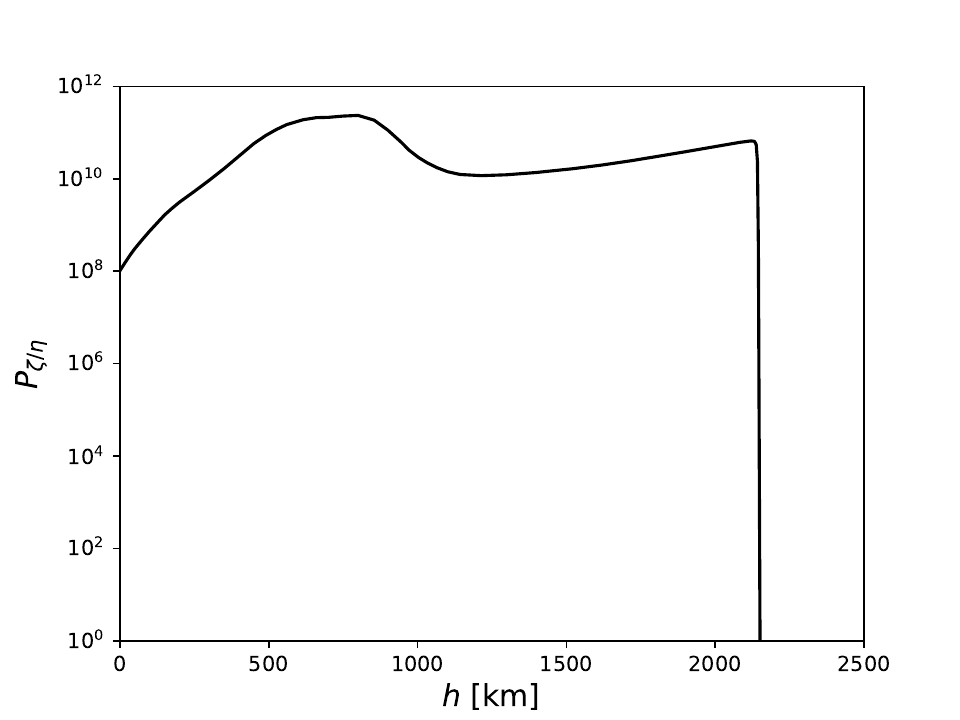}
\caption{The height $h$ profile of bulk viscosity Prandtl number
\Eqref{visc_Prandtl_}
$\mathrm{P}_{\zeta/\eta}\equiv\zeta_0/\eta$ via \citet[Model C7]{Avrett:08}.
One of the purposes of the present study is to emphasize that
for heights $h \equiv x <2.1$~Mm from the solar photospere the bulk viscosity indispensable must be included in the considerations of acoustic wave heating of chromosphere.
For some heights viscous Prandtl number 
$\mathrm{P}_{\zeta/\eta}$ can reach $10^{11}$;
neglecting of $\zeta$ on the background of $\eta$ in this case
is equivalent of neglecting of a giraffe on the background of a hydrogen atom.
}  
\label{fig:prandtl}
\end{figure}

The comparison between different sound waves damping mechanisms
represented in Ref.~\cite[Fig.~1]{PoP:20} for the solar atmosphere via AL08 reveals
that the bulk viscosity is many orders of magnitude larger than the shear one in \Fref{fig:prandtl} where the height dependence of bulk viscosity Prandtl number
\be
\mathrm{P}_{\zeta/\eta}\equiv\frac{\zeta_0}{\eta}\gg 1
\label{visc_Prandtl_}
\ee
is shown.
The collisions of protons with protons 
are more intensive than the collisions with neutral atoms and for the shear
viscosity we use the result for completely ionized hydrogen plasma
\cite[Eq.~(43.9)]{LL10}
\begin{align}
&
\eta\approx0.4\dfrac{M^{1/2}T^{5/2}}{e^4 \Lambda},\qquad
\Lambda=\ln\dfrac{Tr_\mathrm{_D}}{e^2},\\
&
e^2=\dfrac{q_e^2}{4\pi\varepsilon_0},\qquad
\dfrac1{r_\mathrm{_D}^2}=2\,\dfrac{4\pi e^2}{T}n_\rho\alpha,
\end{align}
where $q_e$ is the electron charge, and
$r_\mathrm{_D}$ is the Debye radius \cite[Eq.~(78.8)]{LL5}.
Analogously, the electrons are scattered mainly by protons 
and \cite[Eq.~(43.10)]{LL10}
\be
\varkappa=\dfrac{T^{5/2}}{e^4m^{1/2} \Lambda}.
\ee
Both kinetic coefficients $\eta$ and $\varkappa$ have weak density dependence
only through the Coulomb logarithm $\Lambda$.
The total space damping rate of sound (acoustic) waves \cite[Eq.~(79.6)]{LL6}
\be
k^{\prime\prime}(\omega)
=\frac{\omega^2}{2\rho c^3}\left[
\left(\frac43 \eta+\zeta^\prime(\omega)\right)
+\left(\frac{1}{\mathcal C_v}-\frac{1}{\mathcal C_p}\right)\varkappa
\right]
\label{complete_damping}
\ee
depends on shear $\eta$ and frequency dependent bulk viscosity 
$\zeta^\prime(\omega)$,
thermal conductivity and heat capacities per unit mass
at constant volume $\mathcal C_v$ and pressure $\mathcal C_p$.
For the coefficient of heat conductivity damping in \Eqref{complete_damping}
we have an explicit and computable expression
\be
\frac{1}{\mathcal C_v}-\frac{1}{\mathcal C_p}
=\frac{\Delta\mathcal{C}}
{\mathcal C_v\mathcal C_p}.
\ee
The same approach allows the incorporation of magneto-hydrodynamic (MHD) or Alfv\'en waves (AW)~\cite{Alfven:42,Alfven:47} with vertical magnetic field in the same scheme.
For the AW with the dispersion~\cite[Eq.~(69.8)]{LL8}
supposing that magnetic field is vertical and parallel to the wave-vector
\be
\omega_\mathrm{_{AW}}=V_\mathrm{A}k_\mathrm{_{AW}}^{\prime},\qquad
V_\mathrm{A}=\frac{B}{\sqrt{\mu_0\rho}},
\qquad \mu_0=4\pi
\ee
the damping rate \cite[Problem to Sec.~69]{LL8} 
\begin{align}
2k_\mathrm{_{AW}}^{\prime\prime}
& =(\nu_\mathrm{k}+\nu_\mathrm{m})\,\frac{\omega^2}{V_\mathrm{A}^3},
\qquad
\nu_\mathrm{k} \equiv\frac{\eta}{\rho},
\label{AW_damping}
\\
\nu_\mathrm{m} & \equiv 
\varepsilon_0c^2\varrho_{_{\Omega}}
=\frac{c^2}{4\pi}\frac{e^2m^{1/2}\Lambda}{0.6\,T^{3/2}},
\label{nu_m}
\end{align}
where $\varepsilon_0=1/4\pi$, 
$c$ is the light velocity, and $\varrho_{_{\Omega}}$ is the Ohmic
resistivity \cite[Eq.~(43.8-10)]{LL1}.
The comparison of Alfv\'en wave (AW) damping \Eqref{AW_damping}
with acoustic wave damping \Eqref{complete_damping}
shows that in some circumstances partial ionized plasma could operate as a polarizer  
absorbing longitudinal acoustic waves and transmitting transversal AW. 
Both formulae \Eqref{complete_damping} and \Eqref{AW_damping} 
for the wave damping can be derived as a ratio of
heat production (energy dissipation) $Q_\mathrm{wave}$ divided by the 
energy flux of a plane wave with fixed frequency $\omega$
\be
2k^{\prime\prime}=\mathcal{Q}_\mathrm{wave}/q_\mathrm{wave}.
\ee
Even strong magnetic field 
when parallel to wave-vector
has no influence on the dissipative kinetic coefficients
describing wave damping.

\begin{figure}[ht]
\centering
\includegraphics[scale=0.5]{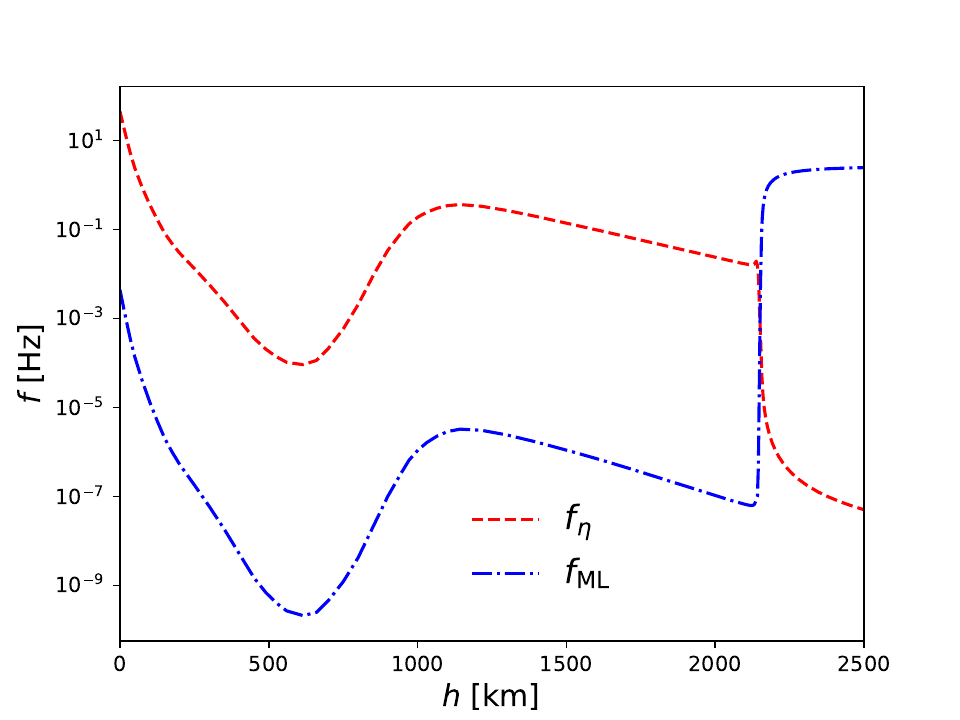}
\caption{Height $h$ profiles of the frequencies 
$f_\mathrm{ML}\equiv 1/2\pi\tau
\ll f \ll 
f_\eta\equiv f_\mathrm{ML}\sqrt{\mathrm{P}_{\zeta/\eta}}$ again via \citet[Model C7]{Avrett:08}.
The damping rate $k_\infty$ is frequency independent 
according to \Eqref{damping_inf}  and determined by the bulk viscosity $\zeta$
and volume heating.
}  
\label{fig:freqs}
\end{figure}

For high enough frequencies the viscosity terms are equalized 
$\zeta_0/(\omega_\eta\tau)^2=\eta$  and above $\omega_\eta \equiv 2 \pi f_\eta$ 
the frequency independent shear viscosity is larger.
In \Fref{fig:freqs} the height $h$ profile of the frequency interval
$(f_\mathrm{ML},\,f_\eta)$ is depicted where the damping rate $k^{\prime\prime}$ is almost frequency independent.
This dramatically simplifies the theoretical analysis; no need to study 
the frequency dependence of the spectral density $\varrho(f)$ of the acoustic waves
\be
q=\int\limits_0^\infty\varrho(f)\md f.
\ee
At the height of the solar transition region $f_\mathrm{ML}$ and $f_\eta$ exchange their places as $\alpha \approx 1$ leading to $\mathcal{D} \rightarrow 0$ in \Fref{fig:D}, meaning that the bulk viscosity heating is abruptly terminated there.

As a concluding remark of this section, it is good to repeat that all formulae derived for the hydrogen-helium cocktail of the solar plasma 
are applicable for laboratory alkaline-noble gas cocktails.

\subsection{Entropy production}

Here we summarize the fundament of the kinetics on which the theory 
of bulk viscosity and wave damping in general is built. 
The main detail of the kinetics is the entropy production and of the hydrodynamics is the heating per unit volume.
Let us recall \cite[Eq.~(49.6)]{LL6}, in the beginning the influence of the shear viscosity and corresponding
incompressible motion of the fluid with velocity vector $\mathbf{v}(t,\mathbf{r})$
\be
Q_\eta=\frac{\eta}{2}\sum_{i,\, j}\left(
\frac{\partial v_i}{\partial x_k}
+\frac{\partial v_k}{\partial x_i}
-\frac{2}{3}\delta_{i,k}\sum_l\frac{\partial v_l}{\partial x_l}
\right)^{\!\!\!2} .
\ee
Analogously for the bulk viscosity heating,
the low frequency formula 
\be
Q_{\zeta_0}=\zeta_0 \left[\mathrm{div}\, \mathbf{v}(t,\mathbf{r})\right]^{2}
\ee
gives
\be
Q_\zeta=\zeta(\omega) 
\left[\Re\!\left(\mathrm{div}\, \mathbf{v}(t,\mathbf{r})\right)\right]^{2},
\ee
where we consider the complex velocity field with fixed frequency $\omega$
\be
\hat{\mathbf{v}}(t,\mathbf{r})=\e^{-\im \omega t}\,\mathbf{v}(\mathbf{r}).
\ee
For the contribution of the temperature gradient for the loss of kinetic energy 
we have
\be
Q_\varkappa=\varkappa(\nabla T)^2,
\ee
and analogously for the ohmic dissipation in magneto-hydro-dynamics
\be
Q_\omega=\rho_\Omega\left(\frac{\mathrm{rot}\mathbf B}{\mu_0}\right)^{\!\!2}.
\ee
Depending on the polarization,
different MHD waves have different
fluid and wave energy dissipation
\be
Q_\mathrm{wave}=Q_\eta+Q_\zeta+Q_\varkappa+Q_{\Omega}.
\ee
Within this scheme, we can take into account the the contribution of 
radiative cooling of the chromospheric plasma as a negative heating
\be
Q_\mathrm{rad}=-\mathcal{P}(T)\, n_e n_p,\qquad n_e=n_p=n_\rho\alpha^2.
\ee
The solar plasma function $\mathcal{P}(T)$ is calculated and tabulated
in great detail in Ref.~\cite{Dere:09}.
For very low temperatures, we can use the linear extrapolation 
in $\ln \mathcal{P}$ versus $1/T$ plot using the lowest temperature $T_0$
for which $\mathcal{P}_0=\mathcal{P}(T_0)$ is tabulated.
In such a way, we obtain the approximation formula
\be
\mathcal{P}(T)\approx \mathcal{P}_0\exp(E_{1,2}(1/T_0-1/T)),\qquad
T<T_0,
\ee
where
\be
E_{1,2}=\frac{I}{1^2}-\frac{I}{2^2}=0.75\, I
\ee
is the difference between $2p$ and $1s$ levels for a hydrogen atom.

Knowing the energy dissipation in the short wavelength approximation, we
can easily calculate the wave damping rate $k^{\prime\prime}(\omega)$,
see different analogous applications for its calculation 
\cite[Eq.~(25.4), Eq.~(79.1)]{LL6}, \cite[Sec.~72]{LL8}.
For the total heating power we have simply
\be
Q=Q_\eta+Q_\zeta+Q_\varkappa+Q_\Omega+Q_\mathrm{rad}
=Q_\mathrm{wave}+Q_\mathrm{rad}.
\ee

Now for illustration, we can write the hydrodynamic equation for the fluid following 
the current lines \cite[Eq.~(5.2)]{LL6}.
Imagine the solar chromosphere in vertical direction on the $x$-axis  and the
static solar wind velocity $U$ also in vertical direction.
In this oversimplified case, the energy flux is \cite[Eq.~(6.3)]{LL6}
\be
q_x=j \tilde{w},\qquad 
\tilde w\equiv\frac12U^2+w+g_\odot x, 
\label{enthalpy_x}
\qquad j\equiv \rho U,
\ee
where $g_\odot$ is the gravitation acceleration. 
For the constant mass flux of the solar wind $j(x)=$ const,
the energy conservation gives the vertical gradient of the generalized enthalpy,
which includes the kinetic and potential energy of the fluid per unit mass
\be
\md_x \tilde w=\frac{Q(x)}{j}.
\label{d_x w}
\ee
For negligible dissipation $Q=0$ we obtain the Bernoulli theorem
\cite[Eq.~(5.1)]{LL6}
\be
\tilde w=\mathrm{const},\qquad s=\mathrm{const},
\ee
but simultaneously at these conditions, the entropy per unit mass $s$
is also constant \cite[Sec.~2, Sec. 5]{LL6}.
For the entropy of a  H-He cocktail
\be
s=\frac{S}{\rho},\qquad S=S_e+S_p+S_0+S_\mathrm{He},
\label{total_s}
\ee
where the $S$ is the additive total volume density of the entropy.
According to the
Sackur-Tetrode formula~\cite{Grimus:13} and  \cite[Eq.~(45.4)]{LL5},
for the partial entropy of ideal gases 
of electrons, protons, neutral hydrogen and helium atoms 
\begin{align}
S_e & =\left[\ln\left(\frac{g_e n_q}{n_e}\right)+c_p\right]n_e,\qquad g_e=2,\,\\
S_p & =\left[\ln\left(\frac{g_p n_{_Q}}{n_e}\right)+c_p\right]n_e,\qquad g_p=2,\,\\
&
n_{_Q} \equiv \left(\frac{MT}{2\pi\hbar^2}\right)^{\! 3/2}=n_q\left(\frac{M}{m}\right)^{\!3/2}, \quad
n_e=n_p, \\
S_0 & =\left[\ln\left(\frac{g_0 n_{_Q}}{n_0}\right)+c_p\right]n_0,\qquad g_0=4,\,\\
S_\mathrm{He} & =\left[\ln\left(\frac{g_\mathrm{_{He}} n_{_A}}{n_\mathrm{_{He}}}\right)+c_p\right]n_\mathrm{_{He}},\qquad g_\mathrm{_{He}}=1,\,
\label{Sackur-Tetrode_He}\\
&
n_{_A} \equiv \left(\frac{AMT}{2\pi\hbar^2}\right)^{\!\! 3/2}=n_{_Q} A^{3/2},\qquad A\approx 4,
\end{align}
where $g_e$, $g_p$, $g_0$ and $g_\mathrm{_{He}}$ are the statistical weights of the particles, 
and $A$ is the mass number of the helium atom,
we obtain
\begin{align}
s=&\frac{1}{M^*} [
-2\alpha \ln\alpha
-(1-\alpha)\ln(1-\alpha)
\label{entropy_explicit} \\
&
-(1+\alpha+\overline a_\mathrm{He})\ln y 
-\overline a_\mathrm{He}\ln \overline a_\mathrm{He}
+(1 + \alpha +\overline a_\mathrm{He})c_p
\nn\\
&
+(2+3\overline a_\mathrm{He})\ln2
+c_v(1+\overline a_\mathrm{He})\ln(M/m) ],\nn
\end{align}
where 
$y\equiv n_\rho/n_q$ 
$=\nu\,\e^{-\iota},$
see \Eqref{nu_equiv}.
Entropy per hydrogen atom (neutral or ionized) 
$s^*\equiv M^*s$ is a convenient dimensionless variable
to represent observable data or theoretical calculations.

\subsection{Entropy of gases and energy quantization}

A nice retrospective review by Grimus~\cite{Grimus:13} on Sackur-Tetrode formula
returns us to the history of the arising of quantum physics of gasses.
On the history of energy quantization, Flamm~\cite{Flamm:97}
presented on the 20-th International Congress of History of Science, 
on July 25$^\mathrm{th}$, 1997, in Li\`ege, Belgium
a review, that we follow.
At the beginning, the main property of quantum mechanics, was perhaps, the potential existence of a discrete energy spectrum and quantum jumps transition between its levels.
In the case of equilibrium, the probability interpretation, together with the combinatorics calculation with the entropy indispensably leads to the quantization of the phase space with an integer number of cells, in contemporary notations $\Delta p \Delta x/(2\pi\hbar)$.
The method of quantization, i.e. introducing of discrete quantum levels of molecules in a finite volume
to obtain the denumerable set of states, was used by Boltzmann in 1872
in the same paper in which he introduced the probability as a dynamic variable to derive an expression for the time derivative of entropy $\md_t S$ explicitly showing that $\md_t S\ge0$~\cite[L. Boltzmann: Weitere Studien\"uber das W\"armegleichgewicht unter Gasmolek\"ulen, Sitzungsber. Kais. Akad. Wiss. Wien Math. Naturwiss. Classe 66 (1872) 275–370] {Boltzmann:72}.
Due to the use of the kinetic equation \Eqref{kinetic_equation}, this diving into the beginning of the story is necessary.
Boltzmann actually introduced cells of finite size in phase-space spanned
on the coordinates and momenta of all molecules. 
Much later, on Born interpreted the Schr\"odinger wave function 
as an amplitude of probability $=\vert\Psi\vert^2$.
In 1877 using quantized energy states in a paper entitled 
``On the relation between the second law of the mechanical theory of heat and the probability calculus with respect to the theorems on thermal equilibrium''
Boltzmann derived the entropy formula 
which in Planck's notation reads $S=\kb \ln W.$
The derivation of the latter finally led to Sackur-Tetrode formula \Eqref{Sackur-Tetrode_He} used in the present paper.
Boltzmann attributed physical significance to the process of energy quantization.
As it was recorded by Ostwald~\cite{Flamm:97,Ostwald:27},
Boltzmann replied to a dispute with Planck at the 64-th meeting
of the Deutscher Naturforscher und \"{A}rzte, Halle, 1891~\cite{Boltzmann:64} with:
\textit{I see no reason why energy shouldn’t also be regarded as divided atomically}, where atomic at that time meant what nowadays is understood as discrete.
This public and recorded event was actually 
the insight moment (Gestalt or illumination) of the birthday of quantum
mechanics (1891, Halle).
Boltzmann suggested to Planck in a paper to apply statistical methods to the problem of blackbody radiation~\cite{Boltzmann:97}.
Oscillator equidistant levels $n=1,\,2,\,3,\,\dots$ and thermodynamic averaging
$\langle n\hbar\omega\rangle=\hbar\omega/[\e^{\hbar\omega/T}-1$]
was the first problem solved, which allowed the comparison with the experimentally measured spectral density of the blackbody radiation.
The Boltzmann method of quantized cells in phase space was explicitly used by Bose in his
representation of the black body radiation.
The Ehrenfest method method for calculation of the 
\textit{Komplextion}~\cite[Eq.~(28)]{Einstein:25,Einstein:25c}
is given as footnote in the Landau-Lifshitz cource on theoretical 
physics~\cite[Sec.~55, Non-equilibrium Fermi and Bose gases]{LL5}:
$\cdot\vert\cdot\cdot\cdot\vert\vert\cdot\cdot\cdot\cdot\vert\cdot\cdot$.
The pictograms explain the method of calculation of 1001 different modes in which 
10 identical Bose particles can be distributed in 5 Boltzmann cells in the phase 
$(\mathbf{p},\mathbf{x})$-space.

The entropy per helium atom with mass $M_\mathrm{_{He}}$
according to Sackur-Tetrode formula \Eqref{Sackur-Tetrode_He}
can be rewritten \cite[Eqs.~(43.5), (45.4)]{LL5} as
\begin{align}
\mathcal{S}_1&\equiv
\frac{S_\mathrm{He}}{n_\mathrm{_{He}}}
\approx-\ln(n_\mathrm{_{He}})+c_v\ln T+(\zeta_\mathrm{_{He}}+c_v+1),\nn\\
\zeta_\mathrm{_{He}}&=c_v\ln\frac{M_\mathrm{_{He}}}{2\pi\hbar^2},\qquad
c_v=\frac32, \qquad T= \kb T^\prime.
\label{chemical_constant}
\end{align}
This low density $n_\mathrm{_{He}}$ approximation formula 
is a simple consequence of the free particle spectrum 
$\varepsilon_\mathbf{p}=p^2/2M_\mathrm{_{He}}$.
However, at unknown temperature in energy units $T$ and
particle mass $M_\mathrm{_{He}}$ this formula did not lead to the
determination of the volume of the Boltzmann cell, i.e. the quantum of the phase
space $2\pi\hbar$ by measuring the chemical constant $\zeta_\mathrm{_{He}}$ .
The method of Boltzmann cells with indistinguishable particles was
applied much later by Bose \cite{Bose:24} for the light spectrum 
$\epsilon_\mathbf{p}=cp$.
The application of this approach for mass particles led in 20-th century
to the creation of the idea of coherent condensation of
Bose particles~\cite{Einstein:25}.

The launching of both ideas of countable states and ``atomic'' structure of the energy
together with their implementation on the equidistant oscillator spectrum, 
were the main events of 19-th century, which triggered the forthcoming development of quantum physics in the 20-th century.

\subsection{Equation for hydrostatic equilibrium}

At differentiation the chemical constant containing $\hbar$ 
disappears and we obtain
in agreement with \Eqref{dp/dT}
\begin{align}
\rho\left(\frac{\partial s}{\partial \rho}\right)_{\!\! T}   
& =-\frac1{M^*} \left[\mathcal{N}_\mathrm{tot}+(c_v+\iota) \mathcal{D}\right]
\label{ds/dro_final} \\
&
= -\frac1{M^*}\!
\left[(1+\alpha+\overline a_\mathrm{He}) +
\frac{(c_v+\iota)\, \alpha(1-\alpha)}{2-\alpha}
\right] \nn\\
&=-\frac1{\rho}
\left(\frac{\partial p}{\partial T}\right)_{\!\! \rho}
=\frac1{\rho}\,\frac{\partial (\varepsilon-sT)}{\partial T\partial \mathcal{V} },
\nn
\end{align}
which is actually the Maxwell relation for the free energy per unit mass
$\varepsilon-sT$.

The vertical component  of the momentum flux of the plasma fluid
\cite[Eq.~(7.2) and Eq.~(7.4)]{LL6}
\be
\tilde p\equiv p+\rho\, U^2=p+\frac{j^2}{\rho}
\label{Pi_x}
\ee
and momentum conservation taking into account gravitation force give
\be
\md_x \tilde p=\md_x\left[p(x)+\frac{j^2}{\rho(x)}\right]=-\rho(x) g_\odot
+\mathcal{F}, 
\label{d_x Pi}
\ee
where $\mathcal{F}$ is the volume density of external for the fluid force 
created by wave absorption.
For example, for a low frequency sound wave propagating in vertical direction
we have in short wavelength approximation of homogeneous fluid
\be
\mathcal{F}\approx\frac{Q_\mathrm{wave}}{c_\infty}, \qquad
Q_\mathrm{wave}\approx 2k^{\prime\prime}(\omega)\,q_\mathrm{wave}.
\ee

For negligible wind velocity and mas and wave energy flux
we obtain the hydrostatic equation \cite[Eq.~(3.1)]{LL6}
\be
\rho(x)=-\frac1{g_\odot}\md_xp.
\label{d_xp}
\ee
The hydrostatics of adiabatic $s(\rho,T)=\mathrm{const}$ 
and iso-enthalpic $w(\rho,T)=\mathrm{const}$
allows tho derive simple equations for this dissipation less motion in homogeneous gravitation field.
The adiabatic condition 
\be
\md_xs=\left(\frac{\partial s}{\partial T}\right)_{\!\!\!\rho}\md_x T
+\left(\frac{\partial s}{\partial \rho}\right)_{\!\!\! T}\md_x \rho=0
\ee
gives
\be
\md_x T=-\dfrac{\left(\dfrac{\partial s}{\partial \rho}\right)_{\!\!T}}
{\left(\dfrac{\partial s}{\partial T}\right)_{\!\!\!\rho}}\,\md_x \rho.
\label{d_xT}
\ee
While the hydrostatic equation \Eqref{d_xp} 
\be
\md_x p=\left(\frac{\partial p}{\partial T}\right)_{\!\!\!\rho}\md_x T
+\left(\frac{\partial p}{\partial \rho}\right)_{\!\! T}\md_x \rho=-g_\odot\rho
\ee
after substitution in $\md_x T$ from \Eqref{d_xT} reads
\be
\md_x\rho=-\dfrac{g_\odot\rho}
{\left(\dfrac{\partial p}{\partial \rho}\right)_{\!\!\! T}
-\dfrac{\left(\dfrac{\partial s}{\partial \rho}\right)_{\!\!T}}
{\left(\dfrac{\partial s}{\partial T}\right)_{\!\!\!\rho}}\,
\left(\dfrac{\partial p}{\partial T}\right)_{\!\!\!\rho}
}.
\label{d_rho/dx}
\ee
Analogously by substituting \Eqref{d_rho/dx} into \Eqref{d_xT},
we obtain
\be
\md_xT=-\dfrac{g_\odot\rho\left(\dfrac{\partial s}{\partial \rho}\right)_{\!\!T}}
{\dfrac{\partial(s,p)}{\partial(T,\rho)}},
\label{d_T/dx}
\ee
where in the denominator here and in \Eqref{d_rho/dx} we have the Jacobian
\be
\frac{\partial(s,p)}{\partial(T,\rho)}=\left(\dfrac{\partial s}{\partial T}\right)_{\!\!\!\rho}\left(\dfrac{\partial p}{\partial \rho}\right)_{\!\!\! T}
-\left(\dfrac{\partial s}{\partial \rho}\right)_{\!\!T}
\left(\dfrac{\partial p}{\partial T}\right)_{\!\!\!\rho}.
\ee
Having an analytical expression for the partial derivatives,
\Eqref{d_rho/dx} and \Eqref{d_xT} give an explicit system of equations 
to calculate the height profiles of the density $\rho(x)$ and the temperature $T(x)$.
The accuracy of the numerical calculation can be evaluated
by verifying whether the generalized enthalpy $\tilde w$ \Eqref{enthalpy_x} 
and simultaneously entropy per unit mass at evanescent solar wind $U(x) \rightarrow 0$ of this profile are constant
\begin{align}
\tilde{w}(\rho(x),T(x))&=\tilde{w}(\rho(x_0),T(x_0))=\mathrm{const}, \\
s(\rho(x),T(x))&=s(\rho(x_0),T(x_0))=\mathrm{const},\\
\tilde w&\equiv\frac12 U^2+\frac{p}{\rho}+\epsilon+g_\odot x.
\end{align}
Instead to solve differential equation,
the adiabatic (and iso-enthalpic) height profile $\rho(x)$ and $T(x)$ can be obtained
by solving the transcedent system of equations above.
Having reliable profiles in dissipation-less approximation we can easily
evaluate the influence dissipation of energy by wave damping considered further
sub-section.


At known wave heating power $Q(x)$ the solution system of equations 
\Eqref{d_x w} and \Eqref{d_x Pi} gives
the height profiles of the temperature $T(x)$ and 
density $\rho(x)$ of the chromospheric plasma.

\subsection{Schwarzschild criterion on convectional instability}

Having analytical expressions for the thermodynamic properties as function of 
$T$ and $\rho$ we can obtain analytical expression for the criterion of  convection
instability \cite{LL6}[Eqs.~(4.4-5)]
\begin{align}
F\equiv
-\frac{\md T}{\md x}<F_c,\;\;\; F_c\equiv\frac{g_\odot\beta T}{\mathcal{C}_p},\;\;\;
\beta\equiv\frac1{\mathcal{V}}\left(\frac{\partial\mathcal{V}}{\partial T}\right)_{\!\!p}.
\label{Schwarz}
\end{align}
As $T= \kb T^\prime$ has dimension of energy, $F_c$ has dimension of force,
and $F_c^\prime=F_c/ \kb $ has dimension of K/m.
For a star $g_\odot=GM_\odot(R)/R^2$.
Using Maxwell relation 
\begin{align}&
\md U=T\md s-p\md\mathcal{V}, \;\;
\md(G\equiv U-Ts+\mathcal{V}p)=-s\md T+\mathcal{V}\md p,\nn\\
&
\frac{\partial G}{\partial p\partial T}=
\left(\frac{\partial\mathcal{V}}{\partial T}\right)_{\!\!p}
=-\left(\frac{\partial s}{\partial p}\right)_{\!\!T}
=-\dfrac{\left(\dfrac{\partial s}{\partial \rho}\right)_{\!\!T}}
            {\left(\dfrac{\partial p}{\partial \rho}\right)_{\!\!T}},
\end{align} 
we obtain for the thermal expansion coefficient
\be
\beta=\frac1{\mathcal{V}}\left(\frac{\partial\mathcal{V}}{\partial T}\right)_{\!\!p}
        =-\dfrac{\rho\left(\dfrac{\partial s}{\partial \rho}\right)_{\!\!T}}
            {\left(\dfrac{\partial p}{\partial \rho}\right)_{\!\!T}}.
             \label{beta}
\ee
Here we have to substitute the derivatives from \Eqref{ds/dro_final}  and \Eqref{dp/d_rho}
and for dimensionless product we finally obtain
\be
T\beta=\dfrac{\mathcal{N}_\mathrm{tot}+(c_v+\iota) \mathcal{D}}
{\mathcal{N}_\mathrm{tot}-\mathcal{D}}.
\label{T_beta}
\ee
As a test when we have completely atomic or completely ionized hydrogen 
we obtain the well known result
\be
T\beta=1,\quad\mbox{for}\; \mathcal{D}=0.
\ee
The substitution $T\beta$ from \Eqref{T_beta} and $\mathcal{C}_p$ 
from \Eqref{C_p} in \Eqref{Schwarz} gives
\be
F_c=\dfrac{ \left[\mathcal{N}_\mathrm{tot}+(c_v+\iota) \mathcal{D}\right]M^*g_\odot}
{\left[c_p\mathcal{N}_\mathrm{tot}
    +(c_vc_p+2c_p\iota+\iota^2)\mathcal{D}\right]\mathcal{N}_\mathrm{tot}},
\label{Schwarzschild_cocktail}
\ee
where we can recognize the averaged mass of the cocktail particles
$M^*/\mathcal{N}_\mathrm{tot}=\langle M\rangle$.
This criterion is applicable under the surface of any star.
The deviation from stability parameter 
$\Delta F\equiv F-F_c$ can be considered as
the driving force of convection.
For negligible influence of the ionization-recombination processes $\mathcal{D}=0$, i.e.
$\alpha=0$ or $\alpha=1$,
the relation above transforms to the well-known
result for the critical temperature gradient 
\be
F_c\approx \langle M\rangle g_\odot/c_p, \qquad c_p=5/2.
\ee
In astrophysics, this formula is applicable under the photosphere where helium is
weakly ionized.
It is an instructive exercise on thermodynamics to check the equivalence of
of both general formulae for the critical temperature gradient 
\Eqref{d_T/dx} and 
\Eqref{Schwarz}
which for our special cocktail gives the explicit result
\Eqref{Schwarzschild_cocktail}.
Simultaneously, this is an analytical test where in a numerical calculation 
all ionized states of the metals are carefully taken into account.
The thermodynamic problem is to check the relation
\be
\mathcal{M}_c(T,\rho)
\equiv-\dfrac{\left.\md_xT\right|_c}{g_\odot}
=\dfrac{\rho\left(\dfrac{\partial s}{\partial \rho}\right)_{\!\!T}}
{\dfrac{\partial(s,p)}{\partial(T,\rho)}}
=\dfrac{\dfrac{T}{\mathcal{V}}\left(\dfrac{\partial\mathcal{V}}{\partial T}\right)_{\!\!p}}{T\left(\dfrac{\partial s}{\partial T}\right)_{\!\!p}},
\ee
where $\mathcal{M}_c$ has dimension of mass and $\rho\mathcal V=1$.
A simple method for solving of the Saha equation in arbitrary atomic plasma 
cocktail is described in \cite{ApJ:24}.
\\
The purpose of this calculation is the intuitive understanding.
Qualitatively, we can analyze the influence of partial ionization 
on convective instability considering simple limit case:
$\alpha=1/2$
and $\mathcal{D}=1/6.$
For cold plasma $\iota\gg1$ we have
\be
F_c\approx\dfrac{M^*g_\odot}
{\left(c_v+\overline{a}_\mathrm{He}\right)}\frac{T}{I}\ll Mg_\odot
\label{qualitative_formula}
\ee
for $ T\ll I\approx158\,\mathrm{kK}.$
In other words, the partial ionization enhance the convective instability of the stars.
Roughly speaking, it is caused by the large heat capacity related to the ionization-recombination processes:
large ionization heat capacity~\cite{ApJ:21}
in the denominator of the critical temperature gradient \Eqref{Schwarz}.  \\
\\
In the general case for arbitrary plasma cocktail, for every chemical element
we calculate the degree of ionization $\overline r_{i,a}$ from \Eqref{Saha_working},
the entropy generalizing $s$ \Eqref{total_s} and pressure $p$
\Eqref{overline_p}.
Their derivatives can be calculated by numerical differentiation in 
\Eqref{C_v_general},
\Eqref{Delta_C},
\Eqref{beta}.
In such a way, we can calculate the convective instability criterion $F_c$ 
\Eqref{Schwarz} for every star with known density and temperature radial profile.\\
\\
Summarizing the analysis of the obtained results here, 
qualitatively small red colder stars are more susceptible to convection 
within their interiors just as it is very well-known.
Low density and low temperature both lead to partial ionization,
provoking the convectional instability.
In the opposite case of massive blue stars, we have only a very narrow 
convective layer.
This is intriguing subject of further and more thorough research.

\subsection{Hydrodynamic equations for stationary wave fluxes}

In order to illustrate the importance of the bulk viscosity in the physics of plasmas,
we consider the simplest possible illustration of a static plasma flux flowing in a pipe with constant cross-section and heated by sound waves.
This could be considered as a very rough model for heating of
the solar chromosphere
by acoustic waves.
We suppose that the generalized enthalpy $w(x)$ \Eqref{enthalpy_x} and 
momentum flux $\Pi(x)$ \Eqref{Pi_x} are static and depend only on the height $x$.
In this case, the energy \Eqref{d_x w} 
and momentum \Eqref{d_x Pi} conservation give the system
\begin{align}
\md_x\tilde w&=\left(\frac{\partial \tilde w}{\partial T}\right)_{\!\! \rho}\md_x T
+\left(\frac{\partial \tilde w}{\partial \rho}\right)_{\!\! T}\md_x \rho=\frac{Q}{j},\\
\md_x\tilde p&=\left(\frac{\partial \tilde p}{\partial T}\right)_{\!\! \rho}\md_x T
+\left(\frac{\partial \tilde p}{\partial \rho}\right)_{\!\! T}\md_x \rho
=\mathcal{F}-\rho g_\odot.
\end{align}
Using that 
\begin{align}
\begin{pmatrix}
\left(\dfrac{\partial \tilde w}{\partial T}\right)_{\!\! \rho}
&\left(\dfrac{\partial \tilde w}{\partial \rho}\right)_{\!\! T}\\
\left(\dfrac{\partial \tilde p}{\partial T}\right)_{\!\! \rho}&
\left(\dfrac{\partial \tilde p}{\partial \rho}\right)_{\!\! T}
\end{pmatrix}
=
\begin{pmatrix}
\left(\dfrac{\partial w}{\partial T}\right)_{\!\! \rho}
&\qquad\left(\dfrac{\partial w}{\partial \rho}\right)_{\!\! T}-\dfrac{U^2}{\rho}\\
\left(\dfrac{\partial p}{\partial T}\right)_{\!\! \rho}&\qquad
\left(\dfrac{\partial p}{\partial \rho}\right)_{\!\! T}-U^2
\end{pmatrix},
\label{grad_tilde_w_p}
\end{align}
after some algebra, we can express explicitly 
the derivatives of the temperature and density
\begin{align}
&
\begin{pmatrix}
\md_x T\\
\md_x \rho
\end{pmatrix}
=\mathsf{W} 
\begin{pmatrix}
-g_\odot+Q/j\\
-g_\odot \rho+\mathcal{F}
\end{pmatrix},
\label{bash_equation}
\end{align}
where
\begin{align}
\mathsf{W} & \equiv
\begin{pmatrix}
\left(\dfrac{\partial \tilde w}{\partial T}\right)_{\!\! \rho}
&\left(\dfrac{\partial \tilde w}{\partial \rho}\right)_{\!\! T}\\
\left(\dfrac{\partial \tilde p}{\partial T}\right)_{\!\! \rho}&
\left(\dfrac{\partial \tilde p}{\partial \rho}\right)_{\!\! T}
\end{pmatrix}
^{\!\!-1}
\label{final_dynamic_system}  \\
&
=
\frac1{\tilde{\mathcal{J}}}
\begin{pmatrix}
\quad \left(\dfrac{\partial p}{\partial \rho}\right)_{\!\! T} -U^2&\qquad
-\left(\dfrac{\partial w}{\partial \rho}\right)_{\!\! T}+\dfrac{U^2}{\rho}\\
-\left(\dfrac{\partial p}{\partial T}\right)_{\!\! \rho} &\qquad
\quad \left(\dfrac{\partial w}{\partial T}\right)_{\!\! \rho}
\end{pmatrix}.
\nn
\end{align}
The determinant in the denominator $\tilde{\mathcal{J}}$ is the Jacobian 
\be
\tilde{\mathcal{J}} \equiv 
\frac{\partial (\tilde{w},\tilde{p})}{\partial (T,\rho)} 
= 
\left(\frac{\partial w}{\partial T}\right)_{\!\! \rho}
\left(\frac{\partial \tilde{p}}{\partial \rho}\right)_{\!\! T}
-\left(\frac{\partial \tilde{w}}{\partial \rho}\right)_{\!\! T}
\left(\frac{\partial p}{\partial T}\right)_{\!\! \rho},
\label{Jacobian_dynamic} 
\ee
which can be rewritten as
\be
\tilde{\mathcal{J}} = \mathcal{J} - U^2
\left [
\left( \frac{\partial w}{\partial T} \right)_{\!\! \mathcal{V}} -
\mathcal{V} \left( \frac{\partial p}{\partial T} \right)_{\!\!\mathcal{V}}
\right ] =
\left(c_0^2-U^2\right) \mathcal{C}_v,
\label{determinant}
\ee
where we have used \cite[Eq.~(16.8)]{LL5} for the enthalpy. 
The $\left(c_0^2-U^2\right)$ factor in the denominator of \Eqref{final_dynamic_system} 
shows that the solar wind cannot reach the sound velocity $U(x)<c_0(x)$ 
for any arbitrary heating function $Q(x)$.
Very high heating leads to $U\rightarrow c_0$ and almost infinite derivatives 
$\md_xT\rightarrow\infty$ and $-\md_x\rho\rightarrow\infty$,
and the latter resembles a standing shock wave.
The numerical solution of these equations will be the subject of a
future hydrodynamic study and in the next section we qualitatively
consider the physical conditions.

The accuracy of the numerical method can be evaluated by the simple test
that at $Q(x)=0$ and $U(x) \rightarrow 0$ we obtain $s(x)=\mathrm{const}.$
Additionally, the solution for the profiles
$T(x)$ and $\rho(x)$ of the system \Eqref{bash_equation} in this case
coincides with the solution of the system \Eqref{d_rho/dx} and \Eqref{d_xT},
or simply stated, 
this dissipation-less approximation is exactly the atmospheric height dependent isoentropic solution.
The profiles $T(x)$ and $\rho(x)$ can be obtained solving the transcendent 
(not differential) system of equations according to \Eqref{entropy_explicit}
and \Eqref{enthalpy_per_unit_mass} completed with \Eqref{alpha_explicit}
\be
w(x)+g_\odot x=w_0\equiv w(0),\qquad
s(x)=s_0\equiv s(0),
\ee
where $w_0$ and $s_0$ are enthalpy and entropy at photosphere surface $x=0$.

\section{A short review of the kinetic coefficients and wave damping in solar plasma}       
                        
\begin{figure}[ht]
\centering
\includegraphics[scale=0.5]{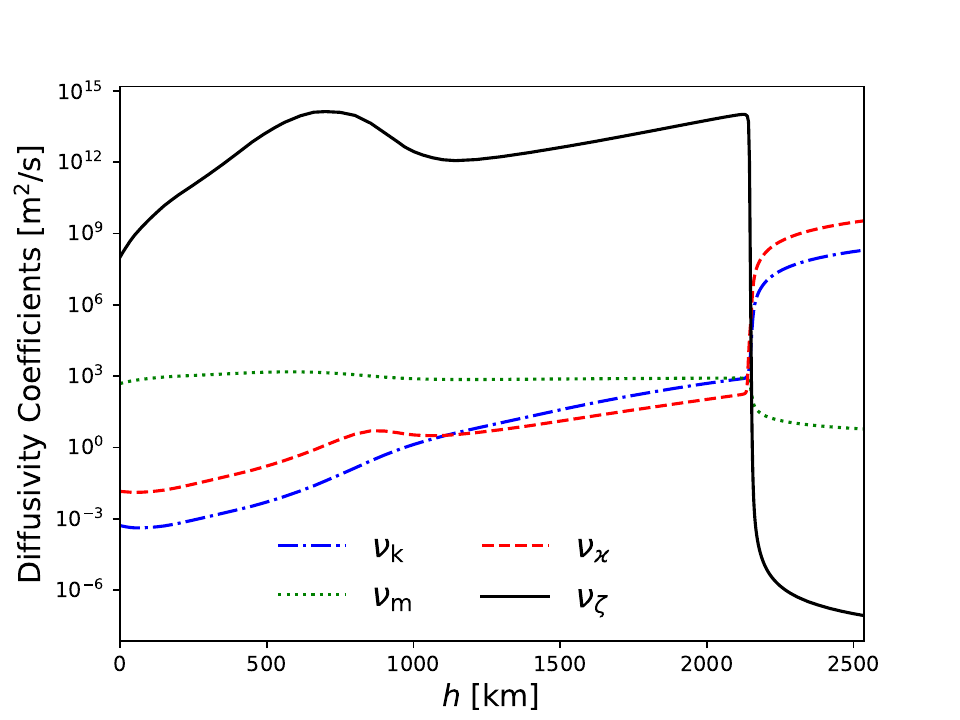}
\caption{Height $h$ profiles of the diffusivity coefficients again via \citet[Model C7]{Avrett:08}.
This figure is very similar to the case for the pure hydrogen plasma cf.~\cite[Fig.~1]{PoP:20}, the differences are in the higher $\nu_\zeta$ and $\nu_\varkappa$ in the current case of H-He
alkali-noble gas cocktail.
The kinematic viscosity $\nu_\mathrm{k}$ is many orders of magnitude 
smaller than the bulk viscosity diffusion coefficient
$\eta/\rho\ll\zeta_0/\rho$ for the whole solar chromosphere.
}  
\label{fig:Diffusion_vs_height}
\end{figure}

Heating of the solar atmosphere by wave damping is actually an old 
idea \cite{Alfven:47}
which undergoes continuous 
evolution, see the recent citations of this classical paper.
The implementation of this idea requires detailed study of the kinetic coefficients
and the bulk viscosity $\zeta_0$, $\tau$ and $\zeta^\prime(\omega)$ \Eqref{Drude_zeta_1} 
was the missing ingredient which was the subject of he present study.
For completeness, we have given the well known for the kinetic coefficients
$\eta$, $\kappa$, and $\varrho_\Omega$
of completely ionized plasma which gives acceptable approximation even for the lower 
chromosphere.
These formulae determine the variables with dimension of diffusivity,
which are important details of many hydrodynamic considerations:
kinematic viscosity $\nu_\mathrm{k}=\eta/\rho$~\Eqref{AW_damping},
$\nu_\zeta=\zeta_0/\rho$,
magnetic diffusivity $\nu_\mathrm{m}=\varepsilon_0 c^2\varrho_{_\Omega}$~\Eqref{nu_m}
and another coefficient related to heat diffusion
\be
\nu_\varkappa=\left(\frac{1}{\mathcal C_v}-\frac{1}{\mathcal C_p}\right)\frac{\varkappa}{\rho}.
\ee
The comparison of these coefficients is extremely instructing for the analysis of 
physical processes in the solar atmosphere and it has already been performed for pure hydrogen in \cite[Fig.~1]{PoP:20}.
The height dependence of all these coefficients is depicted in 
\Fref{fig:Diffusion_vs_height}, where the bulk viscosity $\zeta_0$ and $\nu_\varkappa$ are calculated for the currently studied H-He cocktail, while the other coefficients are calculated for the fully ionized hydrogen plasma
(same as in the cited earlier article).
Comparing both figures, one can note that the He inclusion significantly increases
$\zeta_0$ and slightly increases $\nu_\varkappa$.

We expect that for heating by wave damping the important frequencies are in the mHz range and higher for which $k_\infty^{\prime\prime}$~\Eqref{damping_inf} is dispersion-less,  and 
the sound velocity is $c_\infty$.
At these conditions in the short wavelength approximation the sound energy density
$\propto\e^{-2k_\mathrm{s}^{\prime\prime}}$
where
\be
2k_\mathrm{s}^{\prime\prime}=\tilde a_\mathrm{s} \,\omega^2,\qquad
\tilde a_\mathrm{s}=\left(\frac43\nu_\mathrm{k}+\nu_\varkappa\right)\frac{1}{c_\infty^3}.
\ee
Waves can propagate up to the maximal frequency 
\be
\omega_{c,\mathrm{s}}(x) \equiv \frac1{\tilde a_\mathrm{s} c_\infty}
=\dfrac{c_\infty^2}{\dfrac43\nu_\mathrm{k}+\nu_\varkappa}.
\ee
For Alfv\'en waves we have analogously
\begin{align}
2k_\mathrm{_{AW}}^{\prime\prime}= \tilde a_\mathrm{_{AW}} \,&\omega^2,\quad
\tilde a_\mathrm{_{AW}}=\frac{\nu_\mathrm{k}+\nu_\mathrm{m}}{V_\mathrm{A}^3},\quad
k_\mathrm{_{AW}}^{\prime}=\frac{\omega}{V_\mathrm{A}},\\
\mathrm{for} \quad & \omega \ll \omega_{c,\mathrm{_{AW}}}(x) \equiv
\frac1{\tilde a_\mathrm{_{AW}} V_\mathrm{A}}
=\dfrac{V_\mathrm{A}^2}{\nu_\mathrm{k}+\nu_\mathrm{m}}.
\end{align}
The height dependence of the maximal frequencies are important notions for 
the analysis of the wave damping heating mechanisms of the solar atmosphere.

\begin{figure}[ht]
\centering
\includegraphics[scale=0.5]{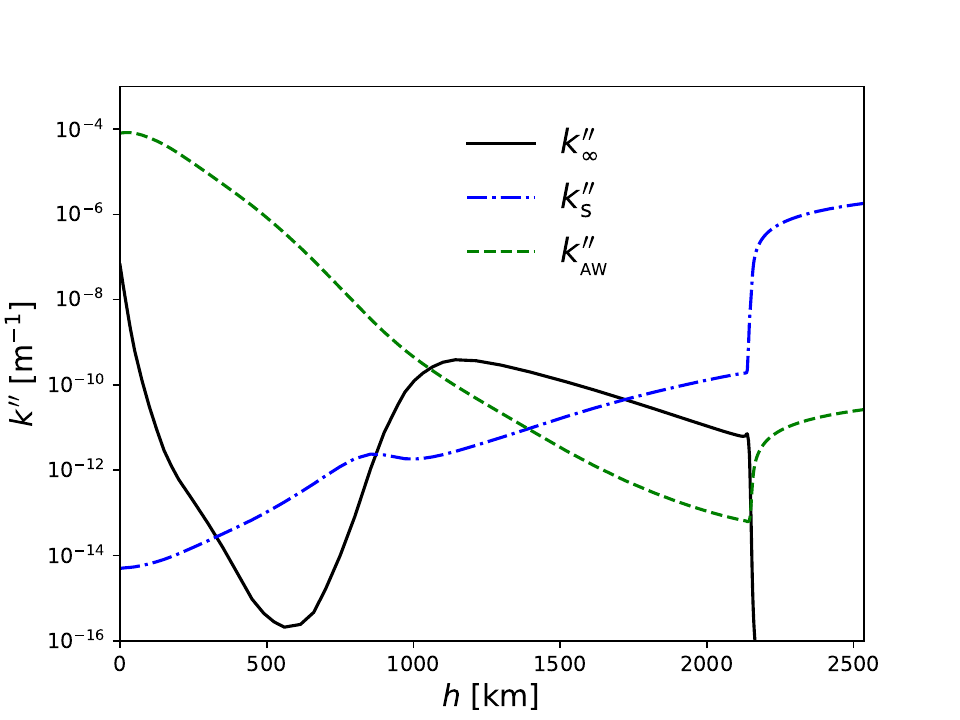}
\caption{Height $h$ profiles of the imaginary parts of the wave-vectors for the diffusivity for the type of different waves for $f = 10$~mHz via \citet[Model C7]{Avrett:08}
($\infty$: sound waves via bulk viscosity, s: sound waves via shear viscosity and heat conductivity and AW: \alf waves via shear viscosity and magnetic diffusivity).
Despite $\nu_\zeta$ being orders of magnitude larger than the rest of the diffusivity coefficients in \Fref{fig:Diffusion_vs_height}, $k_\infty^{\prime\prime}$ dominates in a modest height interval in the solar chromosphere.}  
\label{fig:k_second}
\end{figure}

As a last illustration, the height profiles of the damping rates $k^{\prime\prime}$ for the different damping mechanisms at $f=\rm{10\,mHz}$ are shown in \Fref{fig:k_second}.
One can see that every mechanism has an areal of dominance, which is frequency dependent.
The bulk viscosity damping rate $k_\zeta^\prime\prime$ is frequency independent, but the other two are $\propto \omega^2$, leading to a wider areal of dominance of the bulk viscosity for lower frequencies and a narrower one for higher frequencies respectively.
The influence of the bulk viscosity on the sound waves damping dominates almost everywhere in the solar chromosphere, there is only a narrow region where the influence of the shear viscosity $\eta$ and thermal conductivity $\varkappa$ is significant.
It can be stated that the dominant heating mechanism of the solar chromospheric plasma is sound wave damping by the bulk viscosity.
As a final statement here, all 3 types of damping mechanisms shown here have to be taken into account in studying of the lower solar atmosphere wave dissipation, the AW heating is through the Ohmic resistivity $\varrho$ close to the solar surface and the shear viscosity $\eta$ in the solar transition region, while the sound waves heating is through the bulk viscosity $\zeta$ in between.

\section{Discussion of the applicability of the results 
for acoustic heating of plasmas}

\subsection{Qualitative considerations}
If we evaluate the shear viscosity $\eta$ of the solar chromospheric cocktail which we
use as an illustration,
we observe a huge dimensionless ratio
$\mathrm{P}_{\zeta/\eta}\sim 10^4$ \cite{PhysA}.
For low frequencies $f<f_c$ the bulk (volume) viscosity absolutely
dominates and acoustic heating is created by the ionization-recombination processes.
Let us mention that spectral density of acoustic waves is $\propto 1/f^{\tilde\alpha}$
with $\tilde\alpha\sim 1.6-1.9.$
In the same frequency region however, the absorption of 
\alf waves (AW) is negligible and they are absorbed in the transition region
by the shear viscosity.
The extremely small width of the transition region (TR) is perhaps 
determined by the positive feedback:
AW damping increases the temperature,
increasing the temperature increases the shear viscosity
and wave damping correspondingly, i.e.
a process of self-induced opacity with respect to AW~\cite{First,PHD}.
Below the TR we suppose that the chromosphere is heated
by the bulk viscosity and damping of longitudinal acoustic waves.
Above the TR where these waves have already been absorbed, we have
almost a Naval nozzle created by divergent magnetic force lines
and this nozzle launches the supersonic solar wind.

Perhaps this scenario can be simulated by laboratory plasmas
by a cocktail of alkali and noble gases, Na-Ne cocktail for example.
In the initio of a pipe, magneto-hydrodynamic waves can be inserted in the cold plasma.
The absorption of longitudinal acoustic waves can create
initial heating as in a combustion chamber.
Then as in an after-burn chamber, the transversal AW 
are absorbed in a narrow TR by the 
self-induced opacity.
And finally magnetic field of the enveloping solenoid can simulate
a Naval nozzle for this jet engine for navigation in the solar system.
However, the 3-dimensional problems and possible technical applications 
are far beyond the scope of our task.
In the present work, we have analyzed only the bulk viscosity
and sound wave damping in a homogeneous cold plasma cocktail.

\subsection{Resuming the results}

Let us resume the results:
We have derived the general equations for the ionic concentration oscillations
created by oscillations of the density $\hat\epsilon_{i,a}$
\Eqref{Boltzmann_1} together with oscillation of the temperature
$\hat\epsilon_{_T}$ \Eqref{complex_iota_rho_1}
and electron density $\hat\epsilon_e$ 
\Eqref{electron_density_1}.

These equations are significantly simplified for cold plasmas, where the temperature is much lower than the first ionization potentials of all its ingredients $T\ll I_{a}$;
see \Eqref{Bash_Equation_1} and its matrix representation
\Eqref{epsM_1}.
In order to describe the acoustic heating,
the propagation and absorption of sound waves
with phase velocity $u_\mathrm{s}(\omega)$ and amplitude damping rate 
$k^{\prime\prime}(\omega)$,
we have introduced an useful notion the generalized complex
polytropic index $\hat\gamma(\omega)$
\Eqref{complex_gamma_1}.

For an illustration we use a homogeneous plasma cocktail 
with parameters close to the quiet solar chromosphere.
In this model case, we observe that the
Mandelstam-Leontovich~\cite{Mandelstam:36,Mandelstam:37,Rudenko:77} (ML) approximation perfectly matches the description of frequency dependent bulk viscosity.
We have also revealed how the ML approximation is connected with
Drude~\cite{Drude:00} and Cole-Cole~\cite{ColeCole} approximations invented
phenomenologically for different physical systems.

Returning to the important first ionization rate of the hydrogen atom,
the corresponding cross-section was derived by 
Wannier~\cite{Wannier:53} 
who started in 1953 from the Schroedinger equation.
One can say that the bulk viscosity of cold plasmas $\zeta(\omega)$ is 
one of the few kinetic coefficients allowing first principle calculations.
For pure hydrogen plasma, the problem for the bulk viscosity $\hat\zeta(\omega)$ could have been solved 72 years ago.
Here we repeat the main conclusion of the present study:
for the calculation of height profiles of the temperature and density of the solar atmosphere,
all damping mechanisms have to be taken into account and a midst them the
bulk viscosity of the partially ionized H plasma is an indispensable ingredient.

Source code, input and output datafiles from this study are uploaded in Zenodo~\cite{Zenodo:25}.

\acknowledgments
The authors are thankful to 
Diana Bakkar for the critical reading and significantly improving of the manuscript to
Emil~Petkov for his interest in this study
and to Iglika~Dimitrova for the collaboration in the early stages of the research~\cite{PhysA,ApJ:21}.
The short correspondence with Walter Grimus is also highly appreciated.

\bibliography{zeta}

\end{document}